\documentclass[aps,prb,twocolumn,a4paper,floatfix,showpacs,superscriptaddress]{revtex4}
\usepackage{epsfig}
\usepackage{graphicx}
\usepackage{hyperref}
\usepackage{color}
\usepackage[fleqn]{amsmath}
\usepackage[normalem]{ulem}
\usepackage[switch]{lineno}
\usepackage{multirow}
\usepackage{upgreek} 

\def\HC{hydrogen-induced spot corrosion}
\def\HMC{hydrogen-induced multi-spot corrosion}

\begin{document}

\title{A multiphase-field model for simulating the hydrogen-induced multi-spot corrosion on the surface of polycrystalline metals: Application to uranium metal}

\author{Jie Sheng}
\affiliation{Laboratory of Computational Physics, Institute of Applied Physics and Computational Mathematics, Beijing 100088, China}

\author{Yu Liu}
\email{liu\_yu@iapcm.ac.cn}
\affiliation{Laboratory of Computational Physics, Institute of Applied Physics and Computational Mathematics, Beijing 100088, China}

\author{Xiao-Ming Shi}
\affiliation{School of Materials Science and Engineering,Beijing Institute of Technology, Beijing 100081, China}
\affiliation{Advanced Research Institute of Multidisciplinary Science, Beijing Institute of Technology, Beijing 100081, China}

\author{Yue-Chao Wang}
\affiliation{Laboratory of Computational Physics, Institute of Applied Physics and Computational Mathematics, Beijing 100088, China}

\author{Zi-Hang Chen}
\affiliation{Laboratory of Computational Physics, Institute of Applied Physics and Computational Mathematics, Beijing 100088, China}
\affiliation{School of Materials Science and Engineering,Beijing Institute of Technology, Beijing 100081, China}
\affiliation{Advanced Research Institute of Multidisciplinary Science, Beijing Institute of Technology, Beijing 100081, China}

\author{Ke Xu}
\affiliation{Laboratory of Computational Physics, Institute of Applied Physics and Computational Mathematics, Beijing 100088, China}
\affiliation{School of Materials Science and Engineering,Beijing Institute of Technology, Beijing 100081, China}
\affiliation{Advanced Research Institute of Multidisciplinary Science, Beijing Institute of Technology, Beijing 100081, China}

\author{Shuai Wu}
\affiliation{Laboratory of Computational Physics, Institute of Applied Physics and Computational Mathematics, Beijing 100088, China}
\affiliation{School of Materials Science and Engineering,Beijing Institute of Technology, Beijing 100081, China}
\affiliation{Advanced Research Institute of Multidisciplinary Science, Beijing Institute of Technology, Beijing 100081, China}

\author{Hou-Bing Huang}
\affiliation{School of Materials Science and Engineering,Beijing Institute of Technology, Beijing 100081, China}
\affiliation{Advanced Research Institute of Multidisciplinary Science, Beijing Institute of Technology, Beijing 100081, China}

\author{Bo Sun}
\affiliation{Laboratory of Computational Physics, Institute of Applied Physics and Computational Mathematics, Beijing 100088, China}

\author{Hai-Feng Liu}
\affiliation{Laboratory of Computational Physics, Institute of Applied Physics and Computational Mathematics, Beijing 100088, China}

\author{Hai-Feng Song}
\email{song\_haifeng@iapcm.ac.cn}
\affiliation{Laboratory of Computational Physics, Institute of Applied Physics and Computational Mathematics, Beijing 100088, China}

\date{\today}

\begin{abstract}

Hydrogen-induced multi-spot corrosion on the surface of polycrystalline rare metals is a complex process, which involves the interactions between phases (metal, hydride and oxide), grain orientations, grain boundaries and corrosion spots. To accurately simulate this process and comprehend the underlying physics, a theoretical method is required that includes the following mechanisms: \romannumeral1) hydrogen diffusion, \romannumeral2) phase transformation, \romannumeral3) elastic interactions between phases, especially, the interactions between the oxide film and the hydride, \romannumeral4) elastic interactions between grains, and \romannumeral5) interactions between hydrogen solutes and grain boundaries. In this study, we report a multiphase-field model that incorporates all these requirements, and conduct a comprehensive study of hydrogen-induced spot corrosion on the uranium metal surface, including the investigation of the oxide film, multi-spot corrosion, grain orientation, and grain boundary in the monocrystal, bicrystal, and polycrystal systems. The results indicate that the oxide film can inhibit the growth of hydrides and plays a crucial role in determining the correct morphology of the hydride at the triple junction of phases. The elastic interaction between multiple corrosion spots causes the merging of corrosion spots, and promotes the growth of hydrides. The introduction of grain orientations and grain boundaries results in a variety of intriguing intracrystalline and intergranular hydride morphologies. The model presented here is generally applicable to the {\HMC} on the any rare metal surface.

\end{abstract}
\maketitle
%keywords{Multiphase-field model, Hydrogen-induced spot corrosion, Polycrystals, Uranium hydrides, Hydride formation}

\section{Introduction}
Corrosion refers to the process of loss and destruction of materials through chemical and/or electrochemical reactions with their surrounding media\cite{mccafferty2010introduction}. Most metals resist general corrosion by forming an oxide film (also called a passive film). However, there are some environments where oxide films are not effective in protecting metals, such as pitting corrosion of stainless steel in an electrolyte\cite{szklarska1971review,frankel1998pitting}, or hydrogen corrosion of rare metals, such as cerium\cite{knowles2013morphology,brierley2014microstructure,brierley2015probing}, gadolinium\cite{brill2002effects,benamar2009very,benamar2010heat}, holmium\cite{bloch1982types,bloch2008hydriding}, plutonium \cite{haschke2000surface,brierley2016anisotropic,mcgillivray2011plutonium}, and uranium\cite{owen1966microscope,bingert2004microtextural,hill2013filiform,jones2013surface,stitt2015effects,banos2016effect,appel2018influence,banos2018review,banos2019corrosion1,banos2019corrosion2,ji2019mechanism,banos2020kinetics}. These rare metals are vulnerable to hydrogen attack when exposed to the low levels hydrogen over extended periods. Hydrogen penetrates cracks and voids within the oxide film and react with the internal metal to form the stable metal hydride phase\cite{korst1966rare,mintz1974phase}. Hydrogen corrosion significantly accelerates the degradation of rare metal integrity and durability, ultimately resulting in mechanical failure. Consequently, the investigation of the hydride growth mechanism is particularly important to support the reliable design and accurate assessment of the service life of such metals in hydrogen atmospheres. 

%Especially, uranium and plutonium, as widely used and expensive nuclear engineering materials, are often troubled by hydrogen corrosion,

Hydrogen corrosion on the surfaces of rare metals often manifests as localized attacks resembling ``spots'', known as {\HC}\cite{sheng2022phase}, characterized by four distinct stages: the incubation period, early growth, oxide cracking, oxide spalling, and continued growth. The early growth stage is particularly significant for studying the morphology and kinetics of hydride precipitation, as well as the mechanism behind oxide film deformation and even rupture. Morphological anisotropy\cite{bingert2004microtextural,jones2013surface} and surface bulges\cite{bingert2004microtextural,jones2013surface} are two critical features observed during this stage. These features have garnered significant attention from researchers\cite{bingert2004microtextural,banos2018review,ji2019mechanism}. Previous experimental observations\cite{owen1966microscope,bingert2004microtextural,hill2013filiform,jones2013surface,stitt2015effects,banos2016effect,appel2018influence,banos2018review,banos2019corrosion1,banos2019corrosion2,ji2019mechanism,banos2020kinetics} have indicated that the formation of these features involves hydrogen diffusion, phase transformation, as well as elastic interactions among the oxide film, metal matrix, and hydride precipitate, internal stresses in the material, and interactions between hydrogen solutes and grain boundaries. However, solely relying on experiments presents challenging in revealing the dynamic effects of these diverse factors and capturing the mesoscopic evolution of {\HC} on metal surfaces. Therefore, to advance the research on {\HC} morphology and kinetics, an indispensable tool is a mesoscopic numerical model that effectively couples these factors and accurately predicts the evolution of {\HC}.

Considerable efforts have been made in the development of numerical methods for surface localized corrosion. According to the characterization of the corrosion interface, these techniques can be categorized into two different groups of methods: the sharp interface methods, including the finite volume methods\cite{scheiner2007stable,scheiner2009finite}, the arbitrary Lagrangian-Eulerian methods\cite{sun2014arbitrary}, the level set methods\cite{sethian1996fast} and the diffuse interface method, represented
by the phase-field (PF) method\cite{steinbach2009phase,ansari2021phase,yang2021explicit}. The PF method could avoid tracking the moving interface by constructing the diffuse interface\cite{yang2021explicit}, and has been actively used because of its thermodynamic consistency and convenient coupling of many different physical effects\cite{chen2002phase,li2017review}. Recently, Mai \textit{et al.}\cite{mai2016phase} proposed a simplified PF model for pitting corrosion in steel, capable of reproducing various corrosion kinetics associated with activation-controlled, diffusion-controlled, and mixed-controlled corrosion kinetics. Building upon this work, Ansari \textit{et al.}\cite{ansari2018phase,ansari2019modeling} further extended the PF model by considering electrode reactions to pitting corrosion in steel. Mai and Soghrati\cite{mai2018new}, Chadwick \textit{et al.}\cite{chadwick2018numerical}, and Tsuyuki \textit{et al.}\cite{tsuyuki2018phase} developed the PF model incorporating the effects of ionic species concentration in the electrolyte and electric field strength. These models show promise in simulating mass transport in {\HC}. However, in the {\HC} of rare metals, the elastic interactions between phases need to be further considered in order to adequately apply these models. To address this objective, our previous work\cite{sheng2022phase} introduced a simplified PF model coupled with elastic strain free energy and hydrogen diffusion to investigate {\HC} on uranium surfaces, successfully capturing anisotropic morphologies and surface bulges. Regrettably, for the sake of simplicity, our previous model did not consider the role of the oxide film as a mechanical barrier between the hydride and the environment. In this regard, one idea given by Yang \textit{et al.} \cite{yang2021hydride} was to introduce the oxide film through the multiphase-field (MPF) model. Although this model successfully replicated the corrosion morphology in the metal cerium, it failed to capture surface bulges due to fixed boundary conditions. Furthermore, most of the reported PF models do not take into account real polycrystalline materials with multiple grain boundaries and multiple orientations. Therefore, there is an urgent need to develop a suitable, effective, and accurate polycrystalline MPF model to simulate {\HC} on the surfaces of polycrystalline metals coated with oxide films. 

%the hydrogen diffusion, phase transformation, elastic interaction between phases, elastic interaction between grains, interactions between hydrogen solutes and grain boundaries and
The present work aims to develop and implement a polycrystalline MPF model for approximating the morphology evolution during the {\HC}. Based on the $\operatorname{\mathit{Kim-Kim-Suzuki}}$ (\textit{KKS}) assumption\cite{kim1999phase,kim2004phase}, we propose a polycrystalline MPF model coupling with oxide film, metal, hydride phases, hydrogen concentration, grain boundaries and grain orientation. To verify the validity of the model, the hydride growth on the surface of monocrystal, bicrystal and polycrystal uranium with an exisiting oxide film is simulated. In the monocrystal case, important factors such as oxide film thickness, multi-spot corrosion, and grain orientation are considered to be the key factors to reconstruct spot corrosion morphologies. In addition, we introduce a grain boundary (GB) into monocrystal to form bicrystal structure, allowing us to investigate the influence of GB and multiple grain orientations on the corrosion morphology and internal stress of the material. Finally, we conduct simulations of complex corrosion morphologies in real polycrystalline materials, considering multiple grain boundaries, various grain orientations, and the occurrence of multi-spot corrosion. 

The remainder of this work is organized as follows. Section \ref{sec2} presents a comprehensive description of the MPF model utilized in the simulations, including the governing equations of the model and the total free energy of the system. Section \ref{sec3} provides a variety of numerical simulation results to validate the accuracy and demonstrate the applicability of the proposed model. Finally, section \ref{sec4} concludes the paper by summarizing the key findings and contributions of this study. 
          
\section{The PF model of the {\HC}} \label{sec2} 
\subsection{Problem description}
\begin{figure}%fig1
	\centering
	\includegraphics[width=0.45\textwidth]{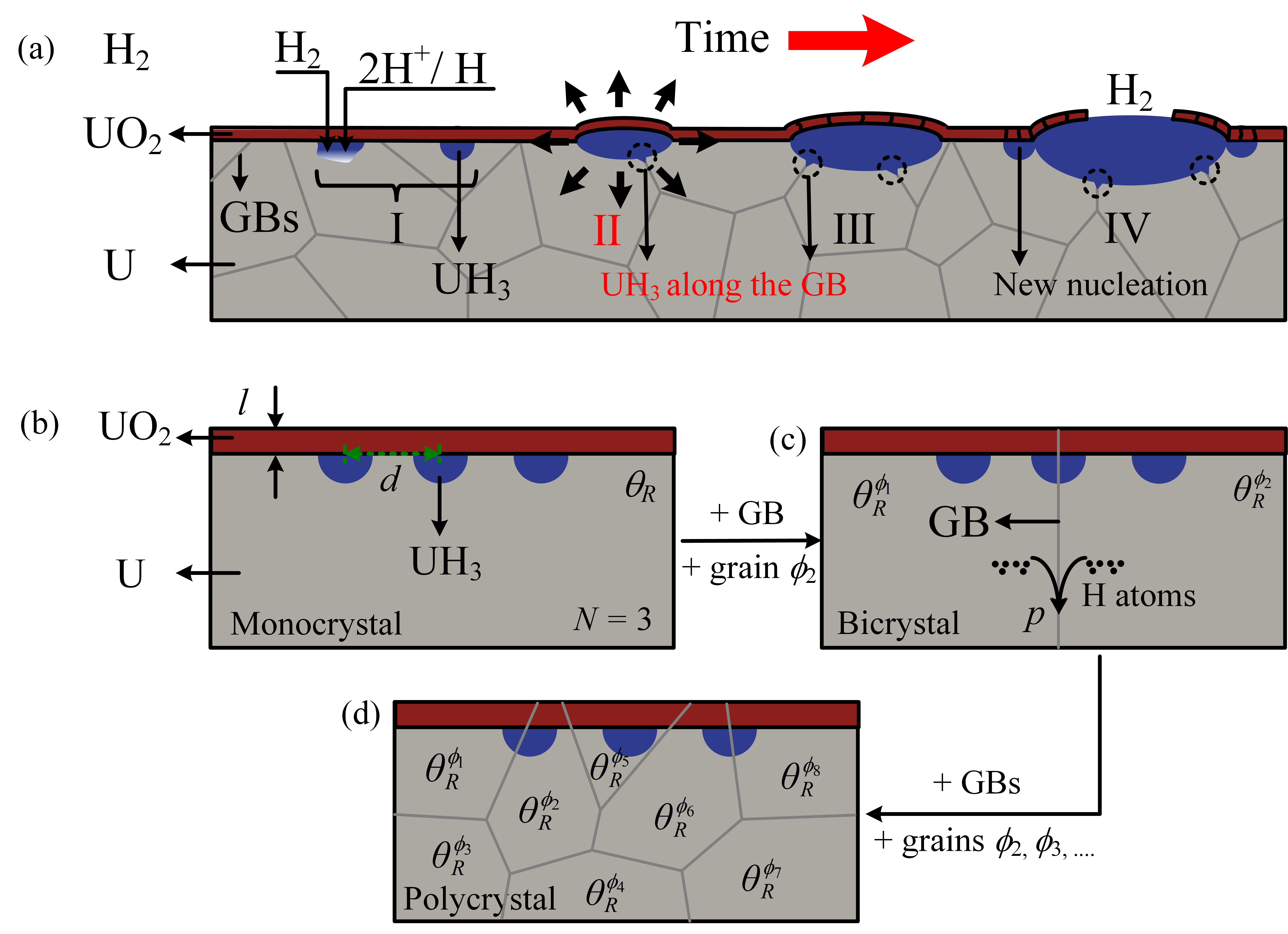}
	\caption{\label{figmodel}(a) Schematic diagram of the {\HC} on the U metal surface\cite{brierley2016anisotropic,banos2018review,sheng2022phase}: \uppercase\expandafter{\romannumeral1}. Incubation period; (\uppercase\expandafter{\romannumeral2}). Early growth; 
		(\uppercase\expandafter{\romannumeral3}). Oxide cracking; 
		(\uppercase\expandafter{\romannumeral4}). Oxide spalling and continued growth. Some key factors affecting corrosion morphology during the (\uppercase\expandafter{\romannumeral2}) stage: (b) Monocrystal: thickness of oxide film $l$, number of corrosion spots $N$, grain orientation $\theta_{R}$; (c) Bicrystal: hydrogen-GB interaction strength $p$, interaction between grain orientation and GB; (d) Polycrystal: all aforementioned factors.}
\end{figure}

Consider a corrosion system\cite{sheng2022phase} consisting of a polycrystalline uranium (U) solid phase coated with an oxide film (mainly UO$_{2}$)\cite{allen1973surface,harker2013altering,banos2018review} in the hydrogen atmosphere, and the resulting corrosion product is $\beta$-UH$_{3}$ at the temperature (513 K) in this work. Fig. \ref{figmodel} illustrates the schematic diagram depicting the process of {\HC} on the surface of U metal in the hydrogen atmosphere\cite{brierley2016anisotropic,banos2018review}. The {\HC} on the metal surface undergoes the following four distinct stages\cite{brierley2016anisotropic,sheng2022phase}:

%Consider a corrosion system composed of a rare metals M solid phase placed in a hydrogen atmosphere.

%i) Incubation period: Under normal conditions, the metal surface is covered with an oxide film, and hydrogen continuously accumulates at the oxide-metal interface to form an initial hemispherical hydride precipitate\cite{owen1966microscope,glascott2003hydrogen,bloch1997kinetics}. ii) Early growth: The initial precipitate grows gradually as the metal and hydrogen react at the metal-hydride interface. In addition, as the hydride precipitate grows, the lattice mismatch between the hydride and the metal creates an outward strain associated with hydride formation, resulting in a bulge on the metal surface\cite{brierley2016anisotropic}. iii) Oxide cracking: the outward strain is sufficient to tensile and distort the oxide film on the surface, causing it to crack. iv) Oxide spalling and continued growth: The oxide film flakes off and the hydride is in direct contact with the environment, accelerating the growth of existing hydride precipitates and the formation of new hydride precipitates.  

(\uppercase\expandafter{\romannumeral1}). Incubation period: Under normal conditions, the metal surface is covered with an oxide film. Hydrogen molecules undergo physical adsorption on the surface of the oxide film, followed by either diffusion or dissociation chemisorption of two H atoms and/or ions\cite{banos2018review}. Subsequently, hydrogen permeates the oxide film in the form of molecules, atoms, or ions, reaching the oxide-metal interface. At the interface, hydrogen accumulates and approaches the hydride solubility limit, leading to the formation of initial hydride nucleation\cite{owen1966microscope,bloch1997kinetics,glascott2003hydrogen}.

(\uppercase\expandafter{\romannumeral2}). Early growth: The initial precipitate gradually grows in size as a result of the reaction between the metal and hydrogen at the metal-hydride interface. This growth process is accompanied by an outward expansion strain caused by the lattice mismatch between the hydride and the metal, resulting in the formation of a bulge on the metal surface\cite{jones2013surface,brierley2016anisotropic}. Additionally, it has been observed that the hydride tends to nucleate and grow at the grain boundaries (GBs)\cite{banos2016effect,brierley2016anisotropic}. 

%An outward expansion strain associated with hydride formation is produced during the growth of hydride precipitate becasue of the lattice mismatch between the hydride and metal\cite{brierley2016anisotropic}, leading to a bulge on the metal surface\cite{brierley2016anisotropic,jones2013surface}. In addition, the hydride preferentially nucleates and grows at the grain boundaries (GBs)\cite{banos2016effect,brierley2016anisotropic}. 
%These hydrides at the GB provides the preferred path for the metal crack growth\cite{brierley2016anisotropic}.

%The oxide film flakes off\cite{jones2013surface} and the hydride is in direct contact with the environment, accelerating the growth of existing hydride precipitates and the formation of new hydride precipitates\cite{brierley2016anisotropic}. 

(\uppercase\expandafter{\romannumeral3}). Oxide cracking: The outward expansion strain caused by the growing hydride precipitate exceeds the tensile strength of the oxide film, resulting in its deformation and subsequent cracking.

(\uppercase\expandafter{\romannumeral4}). Oxide spalling and continued growth: The oxide film on the metal surface undergoes flaking or delamination\cite{jones2013surface}, resulting in direct exposure of the hydride precipitates to the corrosive environment. This direct contact accelerates the growth of existing hydride precipitates and promotes the formation of new hydride precipitates\cite{brierley2016anisotropic}.

%\begin{figure}%
%	\centering
%	\includegraphics[width=0.45\textwidth]{PF2_model_singlecrystal_bicrystal_polycrystal_abstractv1.png}
%	\caption{\label{figsingle_bi_poly_crystal_abstract} Some key factors of the {\HC} in the U metal: (a) monocrystal; (b) bicrystal; (c) polycrystal.}
%\end{figure}

In the present work, our main focus is to investigate the growth of existing hydride precipitates during the early growth stage. Figs. \ref{figmodel}(b)-(c) show some key factors to consider during this stage of the {\HC} in the monocrystal, bicrystal and polycrystal U metals. In the case of the simple monocrystal, our study focuses on examining the morphology and stress of spot corrosion while considering the influence of several key factors. These factors include the thickness of the oxide film ($l$), the number of corrosion spots ($N$), the distance between corrosion spots ($d$), and the effect of grain orientation (represented by the grain rotation angle $\theta_{R}$ with respect to the global coordinate system). Subsequently, we introduce a GB and a grain to form a bicrystal configuration. For the bicrystal, two additional factors come into play: the hydrogen-GB interaction strength ($p$) and the interaction between the grain orientation and the GB. Finally, we extend our analysis to polycrystalline U metal by introducing multiple grains and GBs. By considering all of the above factors, we demonstrate the ability of the MPF model to handle complex grain structures in realistic polycrystalline U metal. 

\subsection{MPF governing equation} \label{sec2.1} 
\subsubsection{General governing equation} \label{sec2.1.1} 
Hydrogen-induced spot corrosion\cite{sheng2022phase} on polycrystalline U with a pre-existing oxide film is a complex process involving five major coupled processes: \romannumeral1) hydrogen diffusion, \romannumeral2) phase transformation, \romannumeral3) elastic interactions between phases, \romannumeral4) elastic interactions between grains, and \romannumeral5) interactions between hydrogen solutes and grain boundaries. To account for and simulate these processes, three types of field variables are required: hydrogen concentration ($c$), phase order parameter ($\eta_\alpha$), and grain order parameter ($\phi_g$). The hydrogen concentration is quantified using the atom fraction for mathematical convenience in describing the hydrogen concentration variation of the system. The phase order parameters require three structural order parameters to characterize the different phases present in the system: the U metal matrix ($\eta_m$), UH$_{3}$ hydride precipitate ($\eta_h$), and UO$_{2}$ oxide film ($\eta_o$) phases. In order to simplify the modeling approach, static grain order parameters are employed. These parameters are necessary for describing the elastic interactions between grains as well as the interactions of concentration with grain boundaries. Note that we use the different indices $\alpha$ (and $\beta, \gamma$) and $g$ for the structural order parameters to identify the phase and grain order parameters in the system, respectively.

%Hydrogen corrosion\cite{sheng2022phase} on the polycrystal U with a pre-existing oxide film involves five major coupled processes: \expandafter{\romannumeral1}) hydrogen diffusion, \expandafter{\romannumeral2}) phase transformation, \expandafter{\romannumeral3}) elastic interactions between phases, \expandafter{\romannumeral4}) elastic interactions between grains, and \expandafter{\romannumeral5}) interactions between hydrogen solutes and grain boundaries. Accordingly, to account for and simulate these processes, three types of field variables are required: hydrogen concentration ($c$), phase order parameter ($\eta_\alpha$), and grain order parameter ($\phi_g$). For simplicity, we adopt static grain order parameters, which are only necessary for describing interactions of concentration with grain boundaries. For the hydrogen concentration, we use the atom fraction for mathematical convenience in describing the hydrogen concentration variation of the system. Note that we use the different indices $\alpha$ (and $\beta, \gamma$) and $g$ for the structural order parameters to identify the phase and grain order parameters in the system. Specifically, the description of the phase order parameters requires three structural order parameters to describe the U metal matrix ($\eta_m$), $\beta$-UH$_{3}$ hydride precipitate ($\eta_h$) and UO$_{2}$ oxide film ($\eta_o$) phases respectively.

In this MPF model, the total free energy functional $F$ of this {\HC} system consists of the interface gradient energy $F_\text{int}$, the bulk free energy $F_\text{bulk}$, the solute-GB interaction energy $F_\mathrm{sg}$ and the elastic energy $F_\text{el}$, which is given by:
\begin{align}\label{eq1}
	F& = F_\mathrm{int} + F_\mathrm{bulk} + F_\mathrm{sg} + F_\mathrm{el} \nonumber\\ 
	&= \int{\left[f_\mathrm{int}(\nabla \boldsymbol{\eta})+ f_\mathrm{bulk}(c,\boldsymbol{\eta}) +f_\mathrm{sg}(c,\boldsymbol{\phi})+f_\mathrm{el}(\boldsymbol{u},\boldsymbol{\eta})\right]}\, dV,
\end{align}
where $\boldsymbol{\eta} = (\eta_m,\eta_h,\eta_o)$ and $\boldsymbol{\phi} = (\phi_1,\phi_2,\phi_3,...)$ being the number of phases and grains in the considered system. $f_\text{int}(\nabla \boldsymbol{\eta})$ is the gradient energy density due to the diffuse interface and $f_\text{bulk}(c,\boldsymbol{\eta})$ is the bulk free energy density depending on the specific materials, $f_\mathrm{sg}(c,\boldsymbol{\phi})$ is the solute-GB interaction energy density, and the elastic energy density $f_\mathrm{el}(\boldsymbol{u},\boldsymbol{\eta})$ is a function of displacement $\boldsymbol{u}$ and phase order parameters $\boldsymbol{\eta}$.

The gradient energy density $ f_\mathrm{int}(\nabla \boldsymbol{\eta}) $ could be formulated as a function of the gradient of the field variables $\eta_\alpha \, (\alpha = m,h,o)$, as follows\cite{villanueva2008multicomponent}: 
\begin{align}\label{eq2}
    f_\mathrm{int} (\nabla \boldsymbol{\eta}) = \sum_{\alpha}{1 \over 2}\kappa_{\alpha}\left| \nabla\eta_\alpha\right|^2, 		
\end{align}
where $\kappa_{\alpha}$ are the gradient energy coefficients for the phase $\eta_\alpha$.

The solute-GB interaction energy density $f_\mathrm{sg}(c,\boldsymbol{\phi})$ denotes the thermodynamically-consistent solute-GB interaction\cite{chen1994computer,heo2011phase,heo2019phase}, which can be written as:
\begin{align}\label{eq3}
f_\mathrm{sg}(c,\boldsymbol{\phi})&= -p \, \omega_g \, c \,q(\boldsymbol{\phi})= - p\, \omega_g \,c\left[\sum_{g}\left(-{1 \over 2}\phi_g^2+{1 \over 4} \phi_g^4\right)\right.  \nonumber \\&  \left. + \sum_{g}\sum_{g^\prime}\phi_g\phi_{g^\prime}+{1 \over 4}\right],		
\end{align}
where $p\,\omega_g$ is the interaction parameter that controls the interaction strength between solute atoms and grain boundaries, and $q(\boldsymbol{\phi})$ is the phenomenological Landau polynomial function of grain order parameters that determines the topology of the solute-GB interaction potential, which is non-zero only at the grain boundaries.

Within the general MPF framework\cite{nestler2005multicomponent,wang2015phase,wang2018phase}, the evolution of the phase order parameter $\eta_\alpha$ is to minimize the free energy functional of the system and can be derived from the functional $F$ via variational derivatives as follows\cite{folch2005quantitative,bollada2012new}:
\begin{align} \label{eq4}
	{\partial \eta_\alpha \over \partial t}& = -L{\delta F \over \delta \eta_\alpha} - \Lambda  \nonumber \\
	&= -L{\delta F \over \delta \eta_\alpha} + {L\over N} \sum_{\beta}{\delta F \over \delta \eta_\beta} \nonumber \\
	&= \sum_{\beta}\left( -\delta_{\alpha\beta}+{1\over N}\right) L\,{\delta F \over \delta \eta_\beta}\nonumber\\
	&= \sum_{\beta}\hat{L}_{\alpha\beta}\frac{\delta F}{\delta \eta_\beta},
\end{align}
where $\Lambda$ is a Lagrange multiplier ensuring the constraint $\sum_{\alpha}\eta_\alpha=1$, and $\Lambda = -{L\over N} \sum_{\beta}{\delta F \over \delta \eta_\beta}$. Here, $\hat{L}_{\alpha\beta} =  \left( -\delta_{\alpha\beta}+{1\over N}\right) L$ is the kinetic coefficient diagonal matrix and $N=3$ in this paper.

\subsubsection{Simplified case (${\partial \eta_o / \partial t} = 0 $)} \label{sec2.1.2} 
The primary focus of this work is to investigate the hydrogenation of U metals coated with a pre-existing oxide film in a hydrogen atmosphere. Due to the absence of oxygen in the environment, metals and hydrides do not transform into new oxides. Therefore, the oxide film phase does not undergo chemical evolution to increase its thickness in the environment. Consequently, the oxide film phase order parameter $\eta_o$ is also a static phase order parameter similar to the grain order parameter. In the future, we will incorporate the effect of oxidation to develop a more comprehensive MPF model that accounts for the oxidation of metal and hydride.

For this static phase order parameter $\eta_o$, its governing equation corresponds to a steady state condition, i.e. ${\partial \eta_o \over \partial t} = 0 $. Substituting the Eq.\eqref{eq4} into the steady state condition, a equation can be derived as follows:
\begin{align} \label{eq5}
 {\delta F \over \delta \eta_o} &= -{1 \over \hat{L}_{oo} }\left( \hat{L}_{om}{\delta F \over \delta \eta_m} + \hat{L}_{oh}{\delta F \over \delta \eta_h}\right) \nonumber \\ &= {1\over 2}\left({\delta F \over \delta \eta_m}+{\delta F \over \delta \eta_h}\right), 
\end{align}
where $\hat{L}_{o\beta} =  \left( -\delta_{o\beta}+{1\over 3}\right) L$  is used in the last step.

Applying the above equation helps to simplify the MPF model, the governing equations of the phase order parameters $\eta_m$ and $\eta_h$ can be rewritten as:
\begin{align}
	{\partial \eta_m \over \partial t} &=\sum_{\beta}\hat{L}_{m\beta}\frac{\delta F}{\delta \eta_\beta}\nonumber \\&=(\hat{L}_{mm}+{1\over 2} \hat{L}_{mo})\frac{\delta F}{\delta \eta_m} + (\hat{L}_{mh}+{1\over 2} \hat{L}_{mo})\frac{\delta F}{\delta \eta_h} \nonumber \\
	&= -{1 \over 2}L\frac{\delta F}{\delta \eta_m}+{1 \over 2}L\frac{\delta F}{\delta \eta_h}, \label{eq6}\\
	{\partial \eta_h \over \partial t} &=\sum_{\beta}\hat{L}_{h\beta}\frac{\delta F}{\delta \eta_\beta}\nonumber \\&=(\hat{L}_{hm}+{1\over 2} \hat{L}_{ho})\frac{\delta F}{\delta \eta_m} + (\hat{L}_{hh}+{1\over 2} \hat{L}_{ho})\frac{\delta F}{\delta \eta_h} \nonumber \\
	&= {1 \over 2}L\frac{\delta F}{\delta \eta_m}-{1 \over 2}L\frac{\delta F}{\delta \eta_h}, \label{eq7}
\end{align}
where $\hat{L}_{\alpha\beta} =  \left( -\delta_{\alpha\beta}+{1\over 3}\right) L$  is used in the derivation. In fact, the governing equation of the phase parameters $\eta_m$ and $\eta_h$ is are equivalent to the conventional PF equation\cite{bollada2012new,sheng2022phase}, but they are expressed in the MPF framework because of the presence of the third phase $\eta_o$ in the system.
    
According to the variational principle ${\delta \over \delta \eta_\alpha} = {\partial \over \partial \eta_\alpha } -\nabla \cdot {\partial \over \partial \nabla \eta_\alpha}$\cite{wang2015phase}, we can derive the variational derivatives of the total free energy $F$ over $\eta_m$ and $\eta_h$, as follows:
\begin{align}
	&\frac{\delta F}{\delta \eta_m}={\partial f_\mathrm{bulk} \over \partial \eta_m} -\kappa_m\nabla^2\eta_m + {\partial f_\mathrm{el} \over \partial \eta_m}, \label{eq8}\\
	&\frac{\delta F}{\delta \eta_h}={\partial f_\mathrm{bulk} \over \partial \eta_h} -\kappa_h\nabla^2\eta_h + {\partial f_\mathrm{el} \over \partial \eta_h}. \label{eq9}
\end{align}

For the governing equation of hydrogen concentration, it could also be derived by minimizing the total free energy $F$ via variational derivatives as the following:

%\begin{align} \label{eq4}
%	{\partial \eta_i \over \partial t} &= -L{\delta F \over \delta \eta_i} + {L\over 3} \sum_{j}^{N}{\delta F \over \delta \eta_j} \nonumber \\
%	&= \sum_{j}\left( -\delta_{ij}+{1\over 3}\right) L\,{\delta F \over \delta \eta_j}\nonumber\\
%	&= \sum_{j}\hat{L}_{ij}\frac{\delta F}{\delta \eta_j}
%\end{align}

\begin{align}\label{eq10}	
	\frac{\partial c}{\partial t} = \nabla\cdot \left( M\nabla\frac{\delta F}{\delta c}\right) = \nabla \cdot \left[ M \nabla \left( {\partial f_\mathrm{bulk} \over \partial c}+{\partial f_\mathrm{sg} \over \partial c}\right)  \right] ,
\end{align}
where $M=D(\boldsymbol{\eta})/\left[ \partial^2 \left( f_\mathrm{bulk}+f_\mathrm{sg}\right) /\partial c^2\right]$ is the mobility. $D(\boldsymbol{\eta})=\sum_{\alpha}D^{\alpha}P_\alpha(\boldsymbol{\eta})$ is the interpolated diffusion coefficient, and $P_\alpha(\boldsymbol{\eta})$ is the interpolated polynomial function. Here, we employed $P_\alpha(\boldsymbol{\eta})= {\eta_\alpha^2 \over 4}\left\lbrace15(1-\eta_\alpha)\left[1+\eta_\alpha-(\eta_\gamma-\eta_\beta)^2\right]+\eta_\alpha(9\eta_\alpha^2-5)\right\rbrace$, where $\alpha,\beta,\gamma$ are all different\cite{folch2005quantitative,choudhury2013quantitative}. These polynomial functions are chosen to satisfy the constraint $\sum_{\alpha}P_\alpha(\boldsymbol{\eta})=1$, ensuring that the total diffusion coefficient accurately represents the contribution from each phase. For the more specific properties of the interpolated polynomial function, the reader is referred to Plapp \textit{et al.}'s work\cite{folch2005quantitative}. 

\subsection{Bulk free energy density}\label{sec2.2} 
For the bulk free energy density, we employ the thermodynamic description of the \textit{KKS} model\cite{kim1999phase,kim2004phase}. In the \textit{KKS} model, each material point is regarded as a mixture of all different phases, and a local equilibrium of chemical potential between all different phases is always satisfied. In this MPF model, these assumptions could be expressed as follows:
\begin{align}
	&c = P_m(\boldsymbol{\eta})c_m+P_h(\boldsymbol{\eta})c_h+P_o(\boldsymbol{\eta})c_o,\label{eq11}\\
	&{\partial f_{m}(c_{m}) \over \partial c_{m}} = {\partial f_{h}(c_{h}) \over \partial c_{h}} = {\partial f_{o}(c_{o}) \over \partial c_{o}},\label{eq12}
\end{align}
where $c_{\alpha}\,(\alpha=m,h,o)$ represent the atom fraction of hydrogen in the $\alpha$-phase, and $f_{m}(c_{m})$, $f_{h}(c_{h})$ and $f_{o}(c_{o})$ in Eq. \eqref{eq12} are the free energy density curves of U metal matrix, UH$_{3}$ hydride precipitate and UO$_{2}$ oxide film phase, respectively. The bulk free energy density could be expressed by a method similar to the concentration assumption in Eq. \eqref{eq11} as follows:
\begin{align}\label{eq13}
	&f_\mathrm{bulk}(c,\boldsymbol{\eta})=\sum_{\alpha}P_\alpha(\boldsymbol{\eta})f_\alpha+\sum_{\alpha}w_\alpha g(\eta_\alpha),
\end{align}
where $w_\alpha$ is the energy barrier parameter for the phase $\eta_\alpha$ and $g(\eta_\alpha)$ is the potential barrier function \cite{folch2005quantitative} for the phase $\eta_\alpha$, which can be formulated as $g(\eta_\alpha) = \eta_\alpha^2(1-\eta_\alpha)^2$.

Following the work of Bair \textit{et al.}\cite{bair2017formation}, the free energy density curves $f_{m}(c_{m})$, $f_{h}(c_{h})$ and $f_{o}(c_{o})$ are constructed based on the parabolic approximation. It is assumed that the free energy density curves $f_{m}(c_{m})$, $f_{h}(c_{h})$ and $f_{o}(c_{o})$ could be written as\cite{sheng2022phase,wu2022phase}: 
\begin{align}
	&f_{m}(c_{m})=A_{m}(c_{m}-c_{1})^2,\label{eq14}\\
	&f_{h}(c_{h})=A_{h}(c_{h}-(c_{2}+\delta c))^2+B_{h},\label{eq15}\\
	&f_{o}(c_{o})=A_{o}(c_{o}-c_{3})^2+B_{o},\label{eq16}\\
	&B_{h} = \Delta G_h/V_{s}, \label{eq17} \\
	&\Delta G_h = 0.1816T-127.67 \rm \left[kJ/mol\right]. \label{eq18} 
\end{align}
Where $c_{1}$ and $c_{3}$ are the concentrations of maximum hydrogen solubility in the U metal matrix and UO$_{2}$ oxide film phases respectively; $B_{o}$ is the decreasing free energy caused by hydrogen dissolution in the UO$_{2}$ oxide film phase. $c_{2}$ is the concentration of hydrogen in the UH$_{3}$ hydride precipitate; $B_{h}$ is the free energy of formation per unit volume of the UH$_{3}$ hydride precipitate; $V_{s}$ is the molar volume of the system\cite{banos2018review} and $\Delta G_h$ is the molar Gibbs free energy of formation of the UH$_{3}$ hydride precipitate\cite{chiotti1980hu}. The $\delta c$ is a small parameter that regulates the magnitude of the thermodynamic driving force\cite{bair2017formation,sheng2022phase}. The constants $A_{m}$, $A_{h}$ and $A_{o}$ in Eqs. \eqref{eq14}, \eqref{eq15} and \eqref{eq16} control the bulk contribution to the interface energy and the tangents between different phases\cite{bair2017formation}. These constants could be determined by constructing the common tangent between the free energy density curves of the UH$_{3}$ hydride precipitate, the UO$_{2}$ oxide film and the U metal matrix (see Sec. \uppercase\expandafter{\romannumeral1} of the Supplementary Materials). A total of two common tangent lines are required: $m-h$, $m-o$ common tangent lines. The parameter $B_{o}$ can also be determined by the $m-o$ common tangent line. 

The maximum hydrogen solubility $c_{1}$ in the U metal matrix phase is usually less than 5 at.\%\cite{hansen1958constitution,massalski1986binary}. For UO$_{2}$ oxide film, its monocrystal has a smaller maximum hydrogen solubility $c_{3}$ (about $0.03$-$0.4\mu$g H$_2$/g UO$_2$)\cite{wheeler1971diffusion} than the U metal matrix, but its internal defects may trap the hydrogen atoms to locally exceed the maximum hydrogen solubility. We assume that the maximum hydrogen solubility of U metal matrix and oxide film are about $c_{1}=3.12$ at.\%\cite{morrell2013uranium} and $c_{3}=1.12$ at.\% in this work. 

\begin{figure}%fig3
	\centering
	\includegraphics[width=0.45\textwidth]{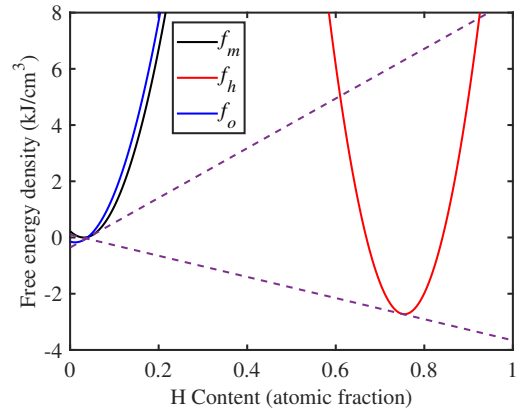}
	\caption{\label{figfree_enrgy}Free energy density curves of U metal (black solid line, $f_{m}$), UH$_{3}$ hydride (red solid line, $f_{h}$) and UO$_{2}$ oxide film (blue solid line, $f_{o}$) in this work; The purple dotted lines represent the phase equilibrium ($m-h$ and $m-o$) common tangents.}
\end{figure}

Fig. \ref{figfree_enrgy} shows the free energy density curves $f_{m}(c_{m})$, $f_{h}(c_{h})$ and $f_{o}(c_{o})$ constructed in this work, where the black solid line corresponds to $f_{m}(c_{m})$, the red solid line corresponds to $f_{h}(c_{h})$, the blue solid line corresponds to $f_{o}(c_{o})$. The dotted line represents the common tangent line. The detailed parameters of the free energy density curves can be found in the Table \ref{tab1}.    

\begin{table}[hb]%table1 %gai chen cheng hao
	\caption{\label{tab1}Model parameters used in the simulations.}
	\begin{tabular}{llll}
		\hline
		Symbol    & Value                                         & Symbol                 & Value                           \\
		\hline
		$T$       & 513 K                                         & $C_{11}^{m}$     & 207.6 GPa\cite{fisher1966temperature} \\                                
		$\kappa_m$ & $1.35 \times 10^{-6}$ J/m                                  & $C_{12}^{m}$	 & 50.6  GPa\cite{fisher1966temperature} \\
		$\kappa_h$ & $1.35\times 10^{-6}$ J/m                                  & $C_{13}^{m}$	 & 24.5  GPa\cite{fisher1966temperature} \\
		$\kappa_o$ & $1.35 \times 10^{-6}$ J/m                                  & $C_{22}^{m}$	 & 185.9 GPa\cite{fisher1966temperature} \\
		$w_m$     & $4.8 \times 10^{10}$ J/m$^{3}$                              & $C_{23}^{m}$     & 102.7 GPa\cite{fisher1966temperature} \\
		$w_h$     & $4.8 \times 10^{10}$ J/m$^{3}$                              & $C_{33}^{m}$     & 242.7 GPa\cite{fisher1966temperature} \\
		$w_o$     & $4.8 \times 10^{10}$ J/m$^{3}$                              & $C_{44}^{m}$     & 108.4 GPa\cite{fisher1966temperature} \\
		$\omega_g$     & $4.8 \times 10^{10}$ J/m$^{3}$                              & $C_{55}^{m}$	 & 55.7 GPa\cite{fisher1966temperature}  \\
		$L$       & $2.17\times 10^{-10}$ m$^{3}$/(J·s)                        & $C_{66}^{m}$ 	 & 63.2 GPa\cite{fisher1966temperature}  \\
		$D^m$     & 0.019{\rm exp}(-5840/\textit{T}) cm$^{2}$/s \cite{powell1973mass} & $C_{11}^{h}$       & 222 GPa\cite{zhang2012electronic}\\
		$D_h$     & 0.019{\rm exp}(-5840/\textit{T}) cm$^{2}$/s \cite{powell1973mass} & $C_{12}^{h}$       & 70  GPa\cite{zhang2012electronic}\\
		$D_o$     & 0.037{\rm exp}(-7199/\textit{T}) cm$^{2}$/s \cite{wheeler1971diffusion} & $C_{44}^{h}$ & 58  GPa\cite{zhang2012electronic}\\
		$A_{m}$   & 235.09 kJ/cm$^{3}$                            & $C_{11}^{o}$	 & 395 GPa\cite{wachtman1965elastic}  \\
		$A_{h}$   & 368.77 kJ/cm$^{3}$           	              & $C_{12}^{o}$     & 121 GPa\cite{wachtman1965elastic}  \\   
		$A_{o}$   & 220.99 kJ/cm$^{3}$           	              & $C_{44}^{o}$     & 64.1 GPa\cite{wachtman1965elastic}  \\   
		$c_{1}$   & 3.12 at.\%\cite{morrell2013uranium}           & $\varepsilon_{11}^{00,o}$	 & 0.0145 \cite{blaxland2015involvement}  \\                     
		$c_{2}$   & 75 at.\%                                      & $\varepsilon_{22}^{00,o}$    & 0.0145\cite{blaxland2015involvement}  \\
		$c_{3}$   & 1.12 at.\%                                    & $\varepsilon_{33}^{00,o}$    & 0.0093\cite{blaxland2015involvement}  \\
		$\delta c$ & 0.5 at.\%                                    & $\varepsilon^{00,h}_{11}$    &0.1259 \\
		$B_{o}$   & -0.1715 kJ/cm$^{3}$                           & $\varepsilon^{00,h}_{22}$    &0.1259 \\
		$V_{s}$   & 12.64 cm$^{3}$/mol\cite{banos2018review}      & $\varepsilon^{00,h}_{33}$    &0.3777 \\
		\hline
	\end{tabular}	
\end{table}

\subsection{Elastic energy density}\label{sec2.3} 
The formation of hydride precipitate within the metal matrix is accompanied by a large volume expansion due to the lower density of the hydride precipitate compared to the metal matrix. The mechanical stresses and strains arising from this volume expansion play a critical role in determining the thermodynamics and kinetics of the phase transformation in {\HC}. To calculate the elastic energy associated with the volumetric changes during hydride formation, we write the elastic energy density based on the Khachaturyan microelasticity theory\cite{morris2010khachaturyan} as the following:
\begin{align}\label{eq19}
	&f_\mathrm{el}(\boldsymbol{u},\boldsymbol{\eta}) = {1 \over 2} \sigma_{ij}\varepsilon_{ij}^{\mathrm{el}}, 		
\end{align}
where $\sigma_{ij}$ and $\varepsilon_{ij}^{\mathrm{el}}$ are stress and elastic strain respectively. The Einstein summation convention is used here and thereafter. Using the additive decomposition theorem for strain under the assumption of small strain and ignoring the plastic strain, the elastic strain is given by:
\begin{align}
	&\varepsilon_{ij}^{\mathrm{el}} = \delta\varepsilon_{ij} - \varepsilon_{ij}^{\mathrm{sf}}, 	\label{eq20}\\
	&\delta\varepsilon_{ij}=\frac{1}{2}\left(\frac{\partial u_i}{\partial r_j}+\frac{\partial u_j}{\partial r_i}\right),\label{eq21}
\end{align}
where $\delta\varepsilon_{ij}$ is the microscopic heterogeneous strain, $\varepsilon_{ij}^{\mathrm{sf}}$ is the stress-free strain associated with the eigenstrain and grain orientation.

To account for the heterogeneous elastic responses in the polycrystalline metal matrix containing the hydride precipitate and oxide film, we adapted the stress-free strain models as the following: 
\begin{align}\label{eq22}  
	&\varepsilon_{ij}^{\mathrm{sf}} = \sum_{g}\zeta_g(\phi_g)R^g_{ik}R^g_{jl}\left( P_h(\boldsymbol{\eta})\varepsilon_{kl}^{00,h}+P_o(\boldsymbol{\eta})\varepsilon_{kl}^{00,o}\right) , 
\end{align}
where $\zeta_g(\phi_g)$ is the grain shape function identifying the $g$-th grain ($\zeta_g = 1$  within the $g$-grain and $\zeta_g = 0$ outside)\cite{heo2019phase}, and $R^g_{ik}$ and $R^g_{jl}$ are the components of a rotation transformation matrix for the rotation from the global reference coordinate system to the local coordinate system defined on the given $g$-th grain \cite{bair2016phase,heo2019phase}. $\varepsilon_{ij}^{00,h}$ and $\varepsilon_{ij}^{00,o}$ are the two kinds of eigenstrains in determining the degree of volume expansion during the phase transformation from the metal to hydride and oxide, respectively. Eq. \eqref{eq22} implies that the eigenstrain exists only in the hydride precipitate and the oxide film phases. The rotation (or orientation) of the $g$-th grain is represented by $\theta^{\phi_{g}}_R$, in two-dimensional (2D) systems, with
\begin{align}\label{eq23}  
	&[{R}^g_{ij}]= \left[ \begin{matrix}
		       \cos \theta^{\phi_{g}}_R  & \sin \theta^{\phi_{g}}_R\\
		       -\sin \theta^{\phi_{g}}_R & \cos \theta^{\phi_{g}}_R \\    
	           \end{matrix}\right]. 
\end{align}

In the Lagrangian finite-strain theory\cite{lubliner2008plasticity}, the eigenstrains $\varepsilon_{ij}^{00,h}$ are related to the deformation gradient tensor by the volumetric strain from the metal matrix to the hydride precipitate. To obtain these vital parameter, we made certain assumptions based on the available experimental data. In the {\HC} experiment of the U metal system, the morphology of hydride precipitate exhibits certain characteristics. Specifically, the hydride precipitates demonstrate approximate isotropy in the $x$-$y$ plane\cite{bloch1997kinetics} and anisotropy in the $z$ direction\cite{jones2013surface}, resembling a round plate shape\cite{bingert2004microtextural,banos2018review}. Moreover, the corrosion depth along the $x$ or $y$ direction is approximately 3 times\cite{blaxland2015involvement} greater than that along the $z$ direction. These observations serve as the basis for our simple assumptions regarding the eigenstrain and deformation gradient tensor. In this work, we assume the eigenstrain $\varepsilon^{00,h}_{ij}$ and the deformation gradient tensors $F_{h}$ as the following:
\begin{align}
	&\varepsilon^{00,h}_{ij}(i\neq j) = 0,\label{eq24} \\
	&\varepsilon^{00,h}_{11} =\varepsilon^{00,h}_{22}=\varepsilon^{00,h},\label{eq25}  \\
    &\varepsilon^{00,h}_{33} =3\,\varepsilon^{00,h},\label{eq26}  \\
	&F_{h} =  \left[ \begin{matrix}
		&1+\varepsilon^{00,h} &0 &0 \\
		&0 &1+\varepsilon^{00,h} &0 \\
		&0 &0 &1+\nu\varepsilon^{00,h}
	\end{matrix}\right], \label{eq27}  
\end{align}
where $\varepsilon^{00,h}$ is a strain parameter that needs to be determined next.

The deformation gradient tensors $F_{h}$ determine the volume expansion and depict the morphological variation in the phase transformation. For this deformation gradient tensor, we should evaluate whether it properly estimates the volumetric strain in order to correctly account for the transformation elastic energy contribution involving the large volume change. Thus, the determinant of $F_{h}$ must satisfy the following equation:
\begin{align}\label{eq28}  
  \mathrm{det}(F_{h})=\Delta V_h,
\end{align}
where $\Delta V_h$ is the volume change of phase transformation from the U metal matrix to UH$_{3}$ hydride precipate, $\Delta V_h \approx 1.746$\cite{banos2018review}. Using the Eq. \eqref{eq27} to solve \eqref{eq28}, we can determine the starin $\varepsilon^{00,h} = 0.1259$. 
%For the parameter $\nu=3$, the strain $\varepsilon^{00,h}$ is 0.1259.

%For the eigenstrain of the oxide film phase, the above method is also applicable, and the volume expansion from metal to oxide is also called the Pilling–Bedworth ratio\cite{lin2020mechanical}. However, for some systems, such as the oxidation of zirconium, there is a large difference between the eigenstrain calculated by the above method and the measured value in the experiment\cite{lin2020mechanical,blaxland2015involvement}. Because the oxide film phase is in contact with the free surface in the $z$ direction, the $z$-direction component of the interal strain can be effectively released by the free surface and the thickening of the oxide film\cite{evans2008influence}. Therefore, the reasonableness of this method in the PF simulation needs to be evaluated\cite{lin2020mechanical}. A similar situation also occurs in the uranium oxidation system. Fortunately, researchers have adopted a more reasonable method to calculate the anisotropic eigenstrain of uranium oxide based on the X-ray diffraction experimental data, which gives $\varepsilon^{00,o}_{11}=\varepsilon^{00,o}_{22}=0.0145$, $\varepsilon^{00,o}_{33}=0.0093$\cite{blaxland2015involvement}.

To evaluate the mechanical response during the phase transformation, we assume that the mechanical equilibrium is established much faster than the diffusion and phase transformations\cite{heo2019phase}, and the displacement is obtained by solving the following mechanical equilibrium equation:
\begin{align}\label{eq29}
	&{\partial \sigma_{ij} \over \partial r_j} = 0.
\end{align}

For simplicity, we consider the linear elastic behavior of the metal matrix, hydride precipitate, and oxide film phases. Therefore, the relationship between stress and elastic strain can be described as follows: 
\begin{align}
	&\sigma_{ij} = C_{ijkl}(\boldsymbol{\eta},\boldsymbol{\phi})\varepsilon_{kl}^{\mathrm{el}},\label{eq30}\\
	&C_{ijkl}(\boldsymbol{\eta},\boldsymbol{\phi}) = \sum_{g}\zeta_g(\phi_g)\sum_{\alpha}P_\alpha(\boldsymbol{\eta})C_{ijkl}^{\alpha}(\phi_g),\label{eq31}\\
	&C_{ijkl}^{\alpha}(\phi_g) = R^g_{im}R^g_{jn}R^g_{ko}R^g_{lp}C_{ijkl}^{\alpha},  \label{eq32}
\end{align}
where $C_{ijkl}(\boldsymbol{\eta},\boldsymbol{\phi})$ is the global elastic constant\cite{heo2019phase} associated with the all phases $\boldsymbol{\eta}$ and all grains $\boldsymbol{\phi}$, $C_{ijkl}^{\alpha}(\phi_g)$ is the elastic constant of the $\alpha$ phase $\,(\alpha=m,h,o)$ in the grain $\phi_g$ and $C_{ijkl}^{\alpha}$ is the reference elastic constants of the $\alpha$ phase in the global coordinate system.

\subsection{Growth model}\label{sec2.4} 
According to the classical nucleation growth theory\cite{porter2021phase}, the growth of the hydride precipitate is governed by the total driving force between the metal and the hydride, which is expressed as\cite{shen2006effect,shen2007effect,han2019phase}:
\begin{align}
	&\Delta G_{m,h} = \Delta G^{\mathrm{ch}}_{m,h} - \Delta G^{\mathrm{el}}_{m,h},\label{eq33}
\end{align}
where $\Delta G^{\mathrm{ch}}_{m,h}$ and $\Delta G^{\mathrm{el}}_{m,h}$ are the chemical driving force and the elastic interaction energy between the metal and the hydride, respectively. 

The chemical driving force can be calculated directly from the vertical distance between the parallel tangents of the respective free energy density curves, as follows\cite{han2019phase,yang2021hydride}:
\begin{align}
	&\Delta G^{\mathrm{ch}}_{m,h} = f_m(c_m)-f_h(c_h)-{\partial f_m(c_m)\over \partial c_m}(c_m-c_h),\label{eq34}
\end{align}
where $f_m(c_m)$ and $f_h(c_h)$ are the free energy density curves of U metal matrix and UH$_{3}$ hydride.

The elastic interaction energy represents the mechanical contribution part of the total driving force, which can be calculated as follows\cite{shen2006effect,shen2007effect,han2019phase}:
\begin{align}
	&\Delta G^{\mathrm{el}}_{m,h} = -\sigma_{ij}\sum_{g}\zeta_g(\phi_g)R^g_{ik}R^g_{jl}\varepsilon^{00,h}_{kl},\label{eq35}
\end{align}
where $\sigma_{ij}$ is the local stress field near the uranium hydride precipitate, $R^g_{ik}$ and $R^g_{jl}$ are the corresponding rotational transformation matrices for the grain $\phi_g$, $\zeta_g(\phi_g)$ is the grain shape function and $\varepsilon^{00,h}_{kl}$ is the egienstrain of UH$_{3}$ hydride precipitate.

Substituting the Eq. \eqref{eq35} to Eq. \eqref{eq33}, we can determine the contribution of elastic interaction to the total driving force. A positive value indicates that the elastic interaction has a negative contribution to the total driving force, suppressing the growth of the hydride precipitate. Conversely, a negative value suggests that the elastic interaction has a positive contribution to the total driving force, promoting the growth of the hydride precipitate\cite{han2019phase}. This relationship between the sign of the elastic interaction and its impact on the total driving force is crucial in understanding the growth behavior of hydride precipitates in the context of {\HC}.

\subsection{Numerical implementation}\label{sec2.5} 
To facilitate the numerical simulation, we employ a nondimensionalized form by defining the characteristic energy ($1.25 \times 10^{-16}$ J), length (5 nm) and time (1 s) for the system. This allows us to express all model parameters in dimensionless units. For solving the evolution equations, the Multiphysics Object Oriented Simulation Environment (MOOSE) based on the Finite Element Method (FEM) is adopted\cite{schwen2017rapid,bair2017formation,lin2019modeling,dykhuis2019phase}. The FEM offers greater flexibility than the finite difference\cite{wheeler1993computation} and spectral methods\cite{li2017review,yang2021hydride}, allowing a wide range of boundary conditions, domain shapes, and coupling to other physics\cite{schwen2017rapid}. It proves to be an effective approach for solving the PF problem with diverse boundary conditions encountered in this study.

To ensure the accuracy of the numerical simulation and to provide computational efficiency, adaptive meshing is utilized to ensure a fine mesh at the interfaces and a large mesh within the phases. In the monocrystal and bicrystal simulations, the interface region contains at least 11 square elements with a size of approximately 2.6 nm to properly approximate the phase order parameters and hydrogen concentration, and the largest mesh size within phases is around 42 nm. For polycrystal simulations, the adaptive mesh adopts the square elements with the smallest mesh size of about 0.65 nm inside the interface and the largest mesh size of about 5 nm inside the phases. 

%To control the adaptive meshing, the gradients of the phase order parameters and the hydrogen concentration are used as refinement criteria and evaluated at each time step. The highest 75\% of the gradient is refined up and the lowest 10\% of the gradient is coarsened up to the specified limits. MOOSE performs the refining or coarsing for mesh based on the Libmesh, the algorithm is described in detail by Stogner et al\cite{stogner2008approximation,kirk2006libmesh}. This adaptive meshing algorithm provides the reliable results and maintains computational efficiency. Furthermore, a second order implicit backward difference method and an adaptive time marching scheme with the absolute tolerance of $5\times 10^{-5}$ and the relative tolerance of $10^{-3}$ are employed to further improve computational efficiency while assuring numerical convergence. 
%放附件

\begin{figure*}[htbp]%
	\centering
	\includegraphics[width=1.0\textwidth]{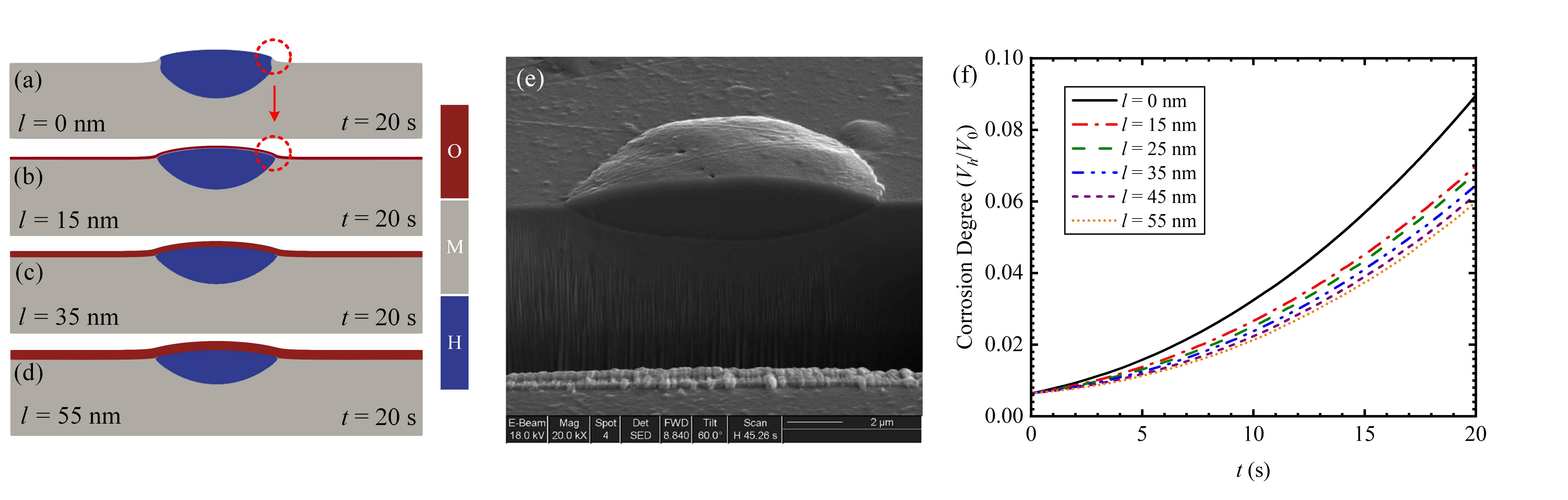}
	\caption{\label{figsingle_thickeness_oxide}Corrosion characteristics under different oxide film thicknesses $l$: (a)-(d) Corrosion morphology at $t=20$ s; (e) Secondary electron images of focused ion beam cross section of corrosion morphology on the surface of U monocrystal; Reproduced with permission from reference\cite{bingert2004microtextural}; (f) Variation of the corrosion degree $(V_{h}/V_0)$ of the metal matrix versus time $t$.}
\end{figure*}

In the concrete simulation, an approach similar to Mai \textit{et al.}'s work\cite{mai2016phase} is adopted. Initially, one or more semicircular initial hydride precipitates with a radius of 70 nm are placed at the top of the rectangular region (0.5 $\upmu$m $\times$ 2.5 $\upmu$m). This region consists of the metal matrix and the oxide film, where the oxide film has a thickness of $l$. The hydrogen concentration in the initial precipitate is set to 75 at.\%, and those of the metal matrix and oxide film are set to 3.12 at.\% and 1.12 at.\%, respectively. Unless otherwise stated, the values of the parameters used for performing the simulations presented in the following sections are selected from Table \ref{tab1}. For the phase field, all boundaries are set to zero flux boundary conditions. For the hydrogen concentration field, the top boundary represents a fixed concentration boundary condition (2 at.\%), simulating the entry of hydrogen from the environment into the phases, while the other boundaries have zero flux boundary conditions. The top boundary of the displacement field is set as free, while the other boundaries are constrained. In the simulation, we define the global reference coordinate system ($x$ and $z$) of the simulation domain along the [100], [001]-directions of the U metal matrix, respectively. The (001) surface of the U metal matrix is exposed to the environment, as previous studies have indicated that the (001) surface is particularly susceptible to hydrogen absorption and {\HC}\cite{taylor2008evaluation,taylor2009ab}. For the bicrystal and polycrystal simulations, the grain rotation angle is used to describe the local coordinate system within each grain. The simulation is carried out under the assumption of plane stress. 

\begin{figure*}[htbp]%
	\centering
	\includegraphics[width=1.0\textwidth]{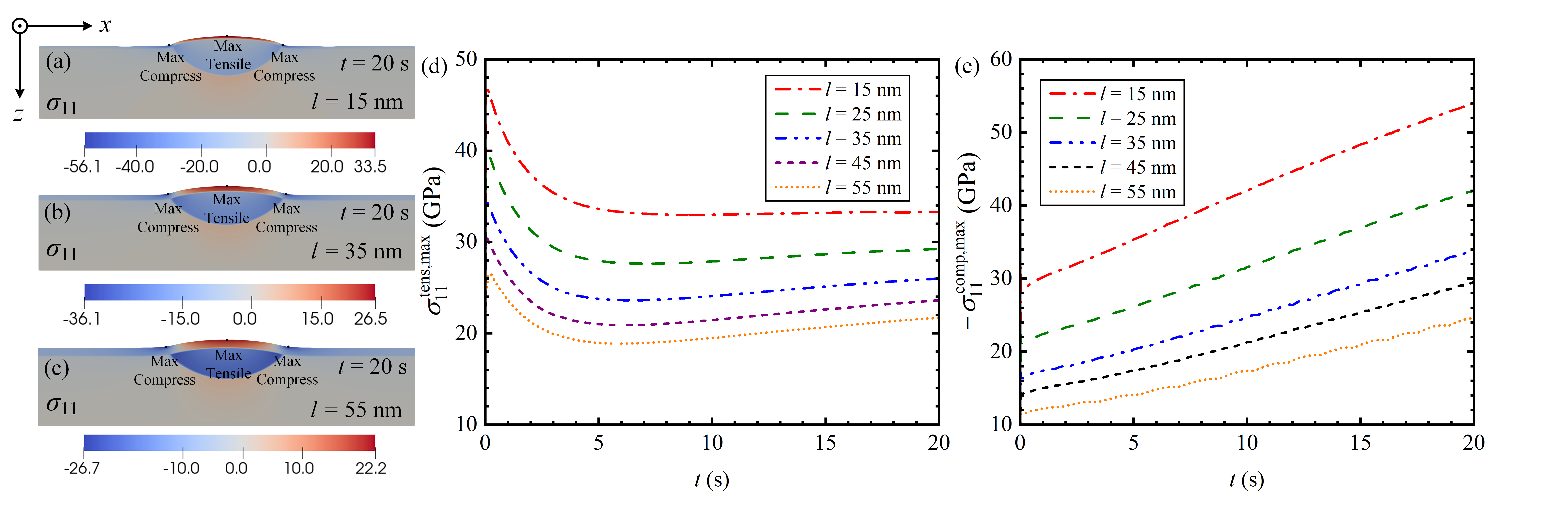}
	\caption{\label{figsingle_stress_oxide}Stress curves during the corrosion process under different oxide film thicknesses $l$: (a)-(c) Distribution of $\sigma_{11}$ (GPa) stress in the metal matrix, hydride precipitate and oxide film at $t=20$ s; Variation of the maximum (d) tensile  ($\sigma^{\mathrm{tens,max}}_{11}>0$) and (e) compressive ($\sigma^{\mathrm{comp,max}}_{11}<0$) stresses versus time $t$ in the oxide film.}
\end{figure*}

\section{Results and Discussion}\label{sec3}  
\subsection{Simulation in the uranium monocrystal} \label{sec3.1} 
A monocrystal system serves as a fundamental and essential model for studying the internal mechanisms of corrosion. Pre-simulation of the {\HC} in the monocrystal can provide valuable insights into the fundamental processes and establish a foundation for investigating bicrystal and polycrystal systems. In this section, we explore the effects of oxide film thickness, the interaction of multiple corrosion spots, and grain orientation on the corrosion morphology and mechanical properties on the surface of U monocrystal. These results validate and demonstrate the accuracy of our MPF model. 

\subsubsection{Effect of oxide film on corrosion morphology} \label{sec3.1.1} 
During the early growth of the hydride precipitate, the oxide film located between the hydride and the external environment acts as a mechanical barrier, impeding the growth of the hydride precipitate\cite{banos2018review}. However, as hydrogen corrosion progresses, the continuous expansion of the hydride could make the failure of the oxide film, leading to the loss of its barrier function\cite{brierley2016anisotropic}. Therefore, it is important to investigate the mechanical interaction between the oxide film and the hydride to accurately estimate the morphology of the hydride and elucidate the fracture mechanism of the oxide film.
 
In the Sec. \uppercase\expandafter{\romannumeral2} A of the Supplementary Materials, we show the evolution of the corrosion morphology under different oxide film thicknesses. As the hydride precipitates expand and grow, the oxide film phase gradually bulges. The hydride precipitate exhibits an anisotropic growth behavior, forming a flat shape. Figs. \ref{figsingle_thickeness_oxide}(a)-(d) show the influence of oxide film thicknesses ($l$) on the flat shape corrosion morphology in the U monocrystal at $t=20$ s. The metal matrix, hydride precipitate, and oxide film phases are represented by the gray, blue, and red colors, respectively. Fig. \ref{figsingle_thickeness_oxide}(a) presents the result simulated using the conventional PF model\cite{sheng2022phase}, while Figs. \ref{figsingle_thickeness_oxide}(b)-(d) show that results obtained using the current MPF model. Increasing the thickness of the oxide film inhibits the longitudinal growth of the hydride precipitate, resulting in a corrosion morphology closer to the plate-like shape. Furthermore, as indicated by the red circles in Figs. \ref{figsingle_thickeness_oxide}(a) and (b), the introduction of the oxide film phase leads to the manifestation of hydride with sharp edges near the triple junction of the phases. These results are consistent with the secondary electron images in the experiments by Bingert \textit{et al.}\cite{bingert2004microtextural} in Fig. \ref{figsingle_thickeness_oxide}(e). Fig. \ref{figsingle_thickeness_oxide}(f) shows the variation of the corrosion degree ($V_h/V_0$) of the metal matrix as a function of time $t$ under various oxide film thicknesses, where $V_h$ represents the volume of the hydride precipitate and $V_0$ is the initial volume of the system. The results demonstrate that increasing the thickness of the oxide film can decelerate the overall corrosion rate of the metal matrix. 

%Fig. \ref{figsingle_thickeness_oxide}(f) depicts the variation of the average elastic energy density $\bar{f}^{\eta_o}_{\rm el}$ of the oxide film phase as a function of time $t$ under different oxide film thicknesses. The average elastic energy density initially decreases as a result of minimizing the elastic free energy through the morphological transformation of hydride from the semicircle to the ellipse (the growth of semi-elliptical corrosion spot is shown in the Sec. \uppercase\expandafter{\romannumeral2} B of the Supplementary Materials). Subsequently, the continuous increase in the average elastic energy density is attributed to the growth of the hydride. In addition, the average elastic energy diminishes as the thickness of the oxide film increases. This can be attributed to the reduction in local strain and stress at material points within the oxide film and the reduction of deformation. In short, a lower average elastic energy stored within the oxide film signifies a smaller volume of hydride located beneath it. These finding highlights the role of the oxide film as a protective barrier against corrosion.

In our previous work\cite{sheng2022phase} on {\HC} of U metals with the absence of an oxide film, we observed significant tensile and compressive stresses $\sigma_{11}$ on the surface of U metals and speculated that these stresses could influence the fracture mechanism of the oxide film. Hence, in this work, we focus on characterizing the stresses within the oxide film. We examine the distribution of stress $\sigma_{11}$ during hydrogen corrosion under different oxide film thicknesses, as depicted in Figs. \ref{figsingle_stress_oxide}(a)-(c). The formed hydride is compressively stressed, which is a consistent result with previous work\cite{heo2019phase,sheng2022phase}. In addition, compressive stress also occurs in the oxide film near the metal interface, while tensile stress occurs in the film above the hydride. These stresses observed within the oxide film can be attributed to two main factors: \romannumeral1) the lattice mismatch between the oxide film and the metal matrix induces a uniform compressive stress in the oxide film\cite{lin2020mechanical} applied by the metal matrix; \romannumeral2) the growth and expansion of hydrides lifts the surface oxide film, applying tensile stress to the center of the oxide film and compressive stress near the edge of the hydride\cite{sheng2022phase}. The maximum tensile stress is concentrated in the center of the oxide film, while the maximum compressive stresses are observed within the oxide film near the hydride edge, indicating a higher risk of fracture in these regions.

Figs. \ref{figsingle_stress_oxide}(d) and (e) illustrate these maximum stresses within the oxide film as a function of time under different oxide film thicknesses. We observe that a decrease in the maximum tensile stress during the initial stages of corrosion, which is associated with the decrease in the elastic energy density within the oxide film induced by the morphological transformation of the hydride (see the Sec. \uppercase\expandafter{\romannumeral2} B of the Supplementary Materials). Subsequently, the maximum tensile and compressive stresses increase continuously due to the continuous growth of the hydride. We also note that increasing the thickness of the oxide film can mitigate the maximum tensile and compressive stresses.

%The morphological transformation of hydride slows down surface bulge and consequently reduces the deformation of oxide film and the maximum tensile stress within the oxide film. 

In short, these simulation results demonstrate the potential of our MPF model to simulate more accurate corrosion morphology and reveal the mechanical interaction between the metal, hydride and oxide film. Moreover, our MPF model effectively captures the stress concentration induced by hydride growth and compression within the oxide film, providing insights into the fracture behavior of the oxide film.

\begin{figure*}[htbp]
	\centering
	\includegraphics[width=1.00\textwidth]{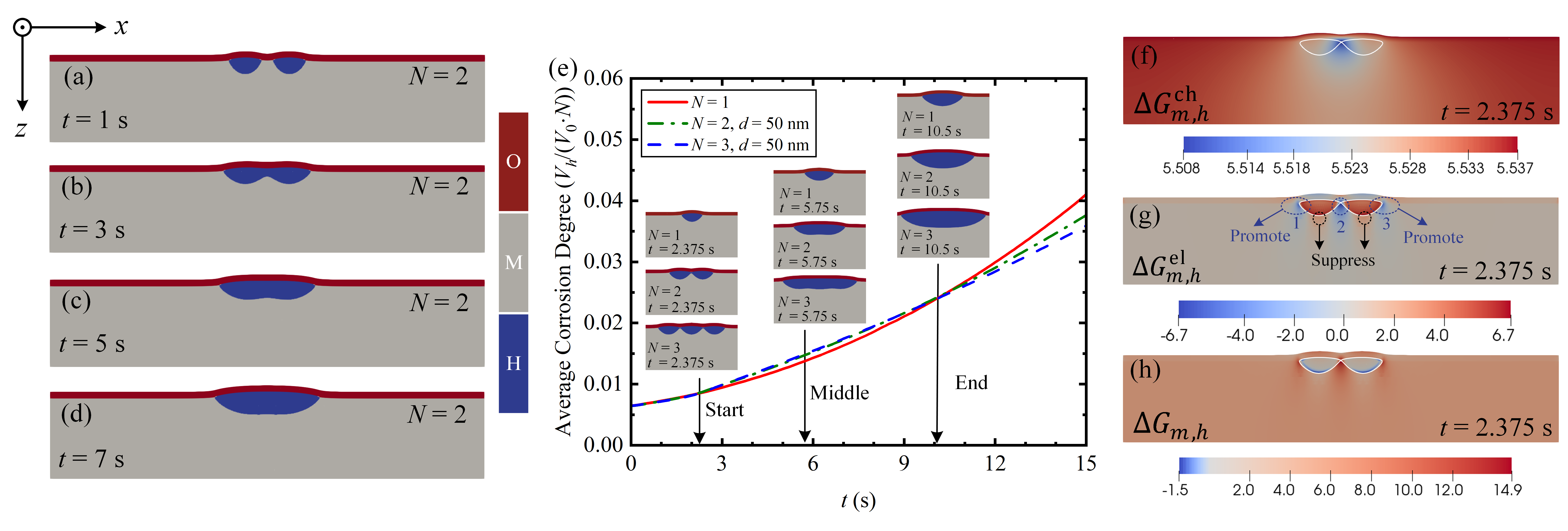}
	\caption{\label{figsingle_multiple_spots}Merging of corrosion spots under different number of corrosion spots $N$: (a)-(d) Evolution of multi-spot corrosion morphology; (e) Effect of corrosion spots merging on the average corrosion degree $(V_h/(V_0 \cdot N))$ of metal matrix. The arrows represents the start, middle and end time of corrosion spots interaction; Distribution of (f) chemical driving force $\Delta G^{\mathrm{ch}}_{m,h}$ (J/mm$^{3}$), (g) elastic interaction energy $\Delta G^{\mathrm{el}}_{m,h}$ (J/mm$^{3}$), and (h) total driving force $\Delta G_{m,h}$ (J/mm$^{3}$) between metal and hydride at the start of corrosion spots interaction ($t = 2.375$ s).}
\end{figure*}

\subsubsection{Effect of multiple corrosion spots on corrosion morphology} \label{sec3.1.2}

The above simulations have focused on the growth of a single corrosion spot, but in reality, several corrosion spots can start adjacent to each other and thus interact with each other as their evolution, forming a larger hydride precipitate. While previous work by Yang \textit{et al.} \cite{yang2021hydride} has explored the influence of spot interaction on corrosion morphology, the underlying mechanism behind the merging of corrosion spots has not been elucidated. In this section, we investigate the effects on corrosion morphology and stress distribution within the material by varying the number of corrosion spots ($N$) and the distance between corrosion spots ($d$).

The evolution of the corrosion morphology in the case of multi-spot corrosion with $N = 2$ is depicted in Figs. \ref{figsingle_multiple_spots}(a)-(d). At the beginning of the simulation, the distance between two identical semicircular corrosion spots is set to 50 nm. At $t = 3$ s, the corrosion spots have interacted and merged, while at $t = 7$ s, the merging of these spots leads to the formation of a larger and more flattened hydride precipitate. The effect of the distance between corrosion spots on the merging of corrosion spots is shown in the Sec. \uppercase\expandafter{\romannumeral2} C of the Supplementary Materials.

\begin{figure*}[htbp]
	\centering
	\includegraphics[width=0.8\textwidth]{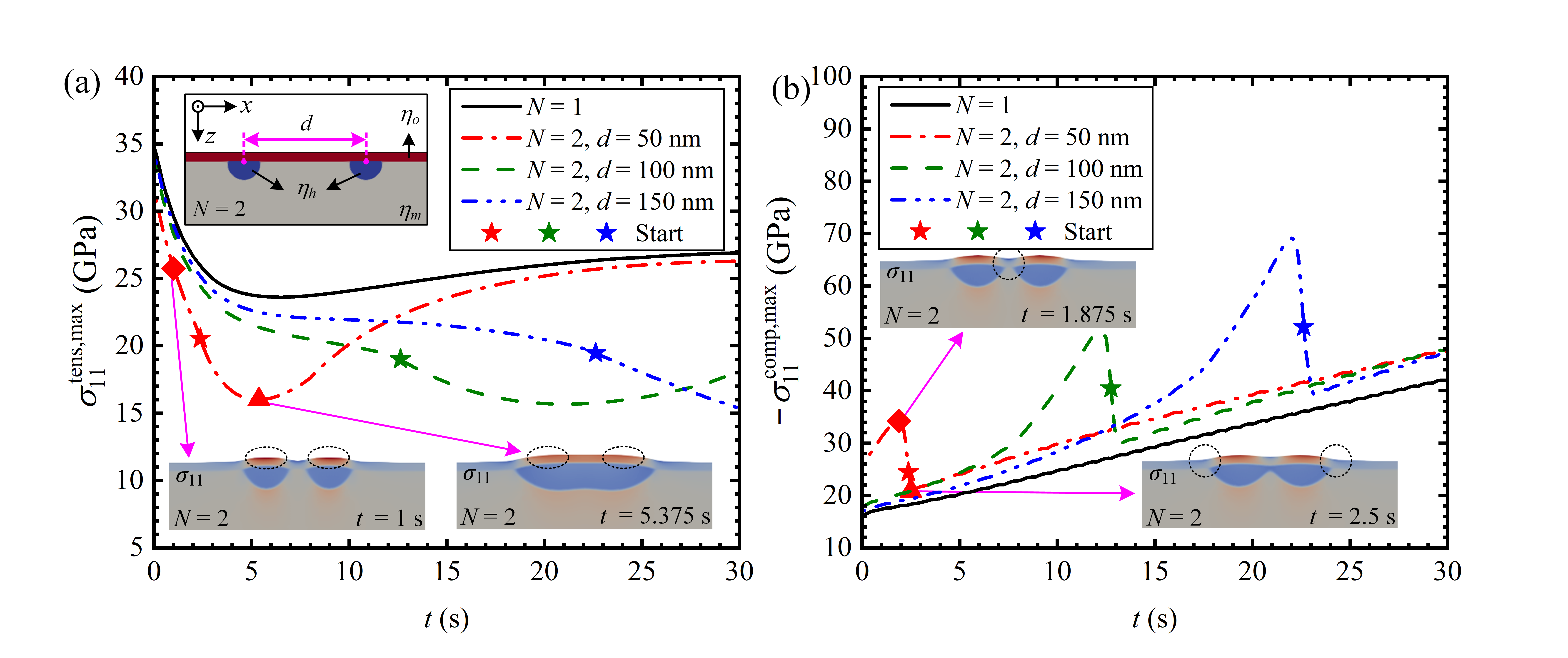}
	\caption{\label{figsingle_multiple_spots_stress}Stress curves during the multi-spots corrosion under different distance $d$: Variation of the maximum (a) tensile  ($\sigma^{\mathrm{tens,max}}_{11}>0$) and (b) compressive ($\sigma^{\mathrm{comp,max}}_{11}<0$) stresses versus time $t$ in the oxide film. The different colored pentagrams indicate the start of corrosion spot interaction. Several insets show the stress distribution of triangle and rhombus points on the stress curve, and the dashed circles in the insets show where the maximum stress is located.}
\end{figure*}

\begin{figure*}[htbp]
	\centering
	\includegraphics[width=1.0\textwidth]{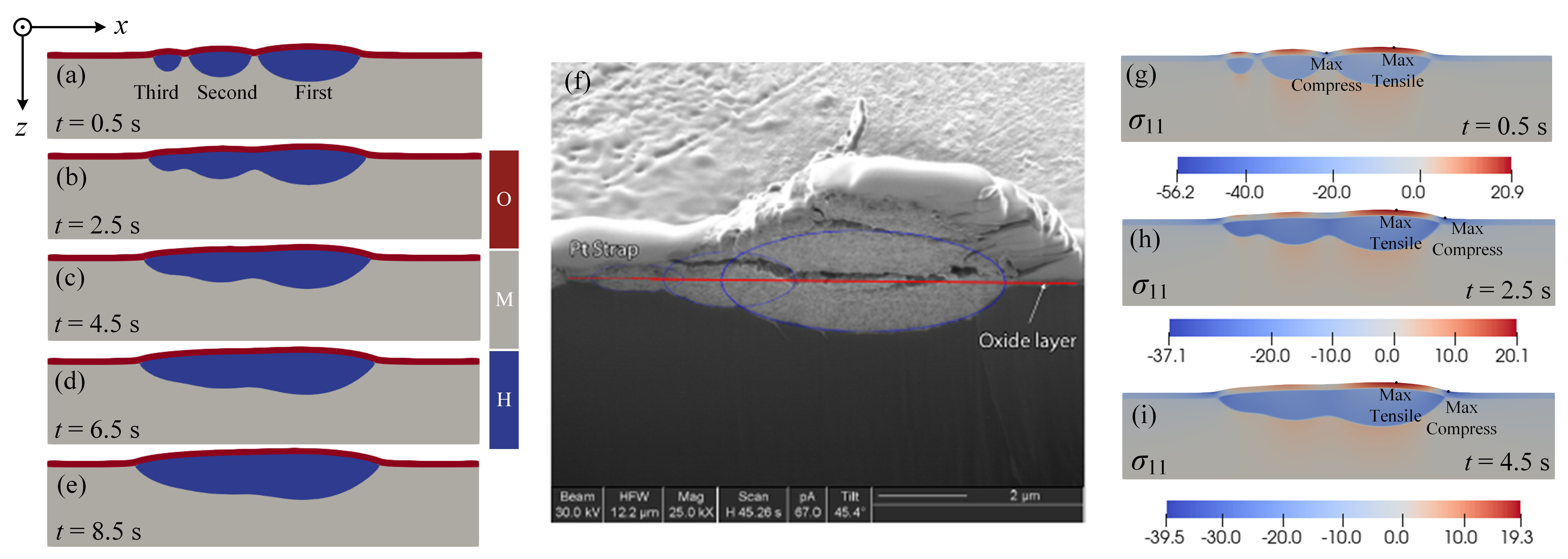}
	\caption{\label{figsingle_multiple_spots_non-simultaneous}Merging of non-simultaneous nucleation corrosion spots($l=35$ nm, $d=50$ nm): (a)-(e) Evolution of non-simultaneous nucleation multi-spots corrosion morphology; (f) Secondary electron images of focused ion beam cross section of asymmetric corrosion morphology on the surface of U monocrystal; Reproduced with permission from reference\cite{jones2013surface} (g)-(i) Distribution of $\sigma_{11}$ (GPa) stress in the metal matrix, hydride precipitate and oxide film.}
\end{figure*}

\begin{figure*}[htbp]%
	\centering
	\includegraphics[width=1.0\textwidth]{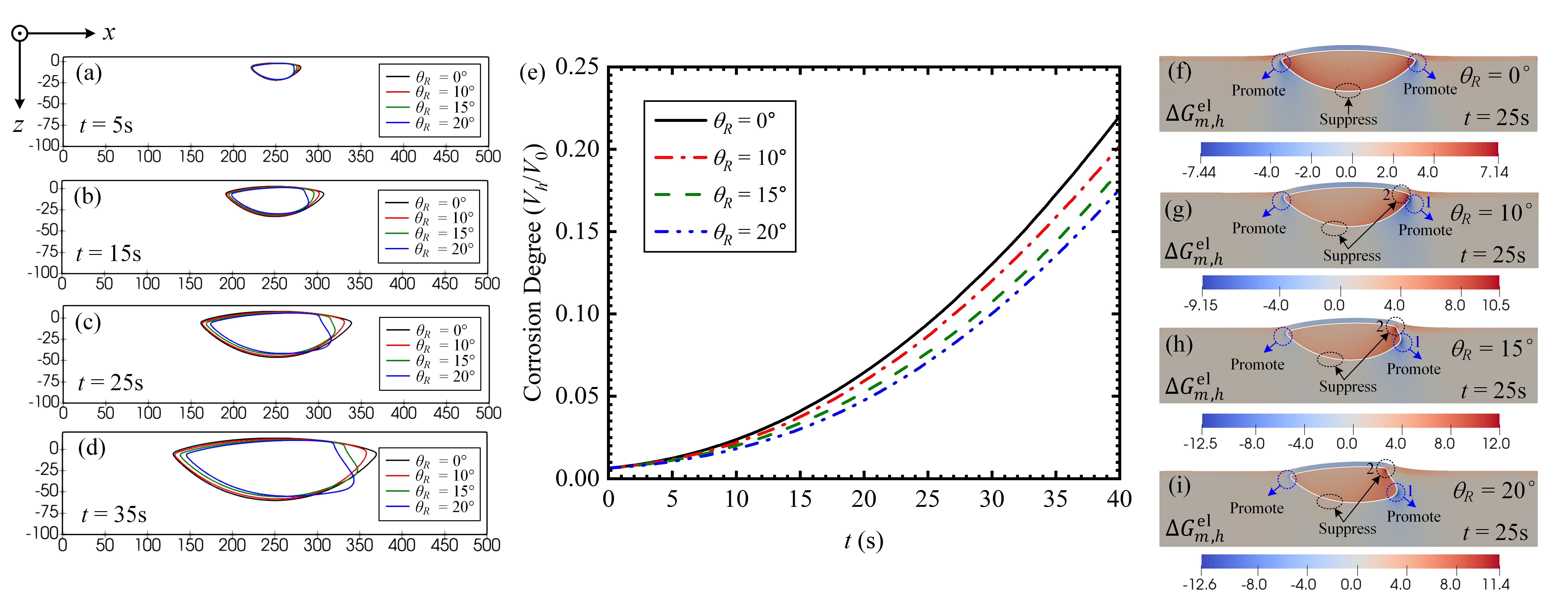}
	\caption{\label{figsingle_orientation_morphology}Corrosion characteristics under different grain orientations $\theta_{R}$: (a)-(d) Propagation of spot corrosion interfaces; (e)Variation of the corrosion degree $(V_{h}/V_0)$ of the metal matrix versus time $t$; (f)-(i) Distribution of elastic interaction energy $\Delta G^{\mathrm{el}}_{m,h}$ (J/mm$^{3}$) between metal and hydride at $t=25$ s.}
\end{figure*}

The influence of the number of corrosion spots on the average corrosion degree $(V_h/(V_0 \cdot N))$ of the metal matrix is shown in Fig. \ref{figsingle_multiple_spots}(e). The insets divide the multi-point corrosion process into three stages. Prior to the interaction between corrosion spots $(t<2.375 \mathrm{s})$, each corrosion spot exhibits a similar average corrosion degree curve for both single-spot and multi-spot corrosion. During the interaction period $(2.375 \mathrm{s}<t<10.5 \mathrm{s})$, the merging of multi-spot corrosion accelerates the average corrosion rate. Subsequently, following the completion of the interaction $(t>10.5 \mathrm{s})$, the average corrosion rate of multi-spot corrosion is lower than that of single-spot corrosion, and the higher the number of corrosion spots, the lower the average corrosion rate. It should be noted that the average corrosion degree solely represents the contribution of each spot to the total corrosion degree $(V_h/V_0)$, and the total corrosion degree for multi-spot corrosion remains significantly higher than that for single-spot corrosion.

In order to elucidate the merging mechanism of corrosion spots and explain its impact on the average corrosion degree, a comprehensive analysis is conducted on the distribution of the chemical driving force $\Delta G^{\mathrm{ch}}_{m,h}$, elastic interaction energy $\Delta G^{\mathrm{el}}_{m,h}$, and total driving force $\Delta G_{m,h}$, as illustrated in Figs. \ref{figsingle_multiple_spots}(f)-(h). Given the uniform distribution of the chemical driving force, the non-uniform elastic interaction energy becomes the dominant factor influencing the corrosion morphology and the merging of corrosion spots. The negative elastic interaction energy (location 2 in Fig. \ref{figsingle_multiple_spots}(g)) between adjacent corrosion spots contributes positively to the total driving force, thereby promoting the merging of corrosion spots and accelerating the average corrosion rate. However, after the merging, the total corrosion degree of multi-point corrosion is less than N times that of single-spot corrosion, resulting in a decrease in the average corrosion rate.

We also investigate the influence of corrosion spot merging on the maximum tensile and compressive stress of the oxide film, as shown in Figs. \ref{figsingle_multiple_spots_stress}(a) and (b), respectively. The different colored pentagrams indicate the start of the corrosion spot interaction. Several insets show the stress distribution of  triangle and rhombus points on the stress curve, and the dashed circles in the insets show where the maximum stress is located. The maximum tensile stress initially decreases due to the shape change of the hydride, further decreases as a result of the merging of corrosion spots, and subsequently exhibits a continuous increase after the merging due to the hydride growth. As the distance between the corrosion spots increases, the corrosion spots require a longer growth time to contact and interact, and the further reduction in the maximum tensile stress caused by the merging of the corrosion spots is delayed accordingly. In addition, the merging of corrosion spots can relieve the compressive stress between them. Prior to the merging of corrosion spots, the maximum compressive stress of the oxide film is concentrated between adjacent corrosion spots (the inset when $t=1.875$ s). However, during the merging process, the maximum compressive stress between adjacent corrosion spots is alleviated and shifts towards the oxide film near the hydride edges (the inset when $t=2.5$ s). Increasing the distance between corrosion spots can also delay the release of compressive stress. 

Interestingly, Figs. \ref{figsingle_multiple_spots_stress}(a) and (b) reveal that the maximum tensile stress and compressive stress within the oxide film for the multi-spot corrosion can revert to the stress levels observed for the single spot corrosion after the merging of corrosion spots. After the merging of the corrosion spots, the maximum compressive stress quickly returns to the stress level seen for the single spot corrosion (compare the red, green, and blue curves with the black curve). In contrast, the recovery of the maximum tensile stress takes a longer duration. This is due to the fact that the hydride needs time to grow sufficiently to produce a larger bulge, ultimately leading to the positional relocation of the maximum tensile stress toward the center of the oxide film. 

%In contrast, the recovery of the maximum tensile stress takes a longer duration as it necessitates a significant amount of time for the morphological changes in the hydride to shift the position of the maximum tensile stress towards the center of the oxide film. This delay is attributed to the time required for the hydride to undergo sufficient growth and redistribution within the corrosion spot, ultimately leading to the positional relocation of the maximum tensile stress. 

Figs. \ref{figsingle_multiple_spots_non-simultaneous}(a)-(e) present a simulation illustrating the merging of non-simultaneous nucleation corrosion spots, resulting in an asymmetric corrosion pattern. The results are in good agreement with the secondary electron images from experiments by Jones \textit{et al.}\cite{jones2013surface} in Fig. \ref{figsingle_multiple_spots_non-simultaneous}(f). This morphology is formed by the merging of three adjacent corrosion spots that nucleated at different times, and the nucleation sequence of the corrosion spots is indicated. 

%In comparison to the merging of multiple simultaneous nucleation corrosion spots above, that of non-simultaneous nucleation corrosion spots results in an asymmetric corrosion pattern.
%Fig. \ref{figsingle_multiple_spots_non-simultaneous}(f) shows the SEM image about observed in experiments\cite{jones2013surface}. The results demonstrate that the simulated asymmetric corrosion morphology of metal are in good agreement with experiments\cite{jones2013surface}.

Figs. \ref{figsingle_multiple_spots_non-simultaneous}(f)-(h) show the distribution of stress $\sigma_{11}$ in the asymmetric corrosion pattern. The maximum tensile stress is consistently located in the center of the oxide film above the largest of the three corrosion spots due to the largest bulge of the oxide film above the largest corrosion spot. However, the maximum compressive stress appears initially between adjacent corrosion spots, and as the corrosion spots merge, it shifts to the oxide film near the location of the largest corrosion spot.

The above results show that the merging of corrosion spots accelerates the corrosion rate. The closer the distance between the corrosion spots, the faster the merger occurs. The merging of multiple adjacent corrosion spots that nucleated at different times can induce the formation of asymmetric morphology. This merging phenomenon highlights the significance of corrosion spot interaction in influencing the overall corrosion morphology, corrosion rate and material stress.

\begin{figure*}[htbp]%
	\centering
	\includegraphics[width=0.8\textwidth]{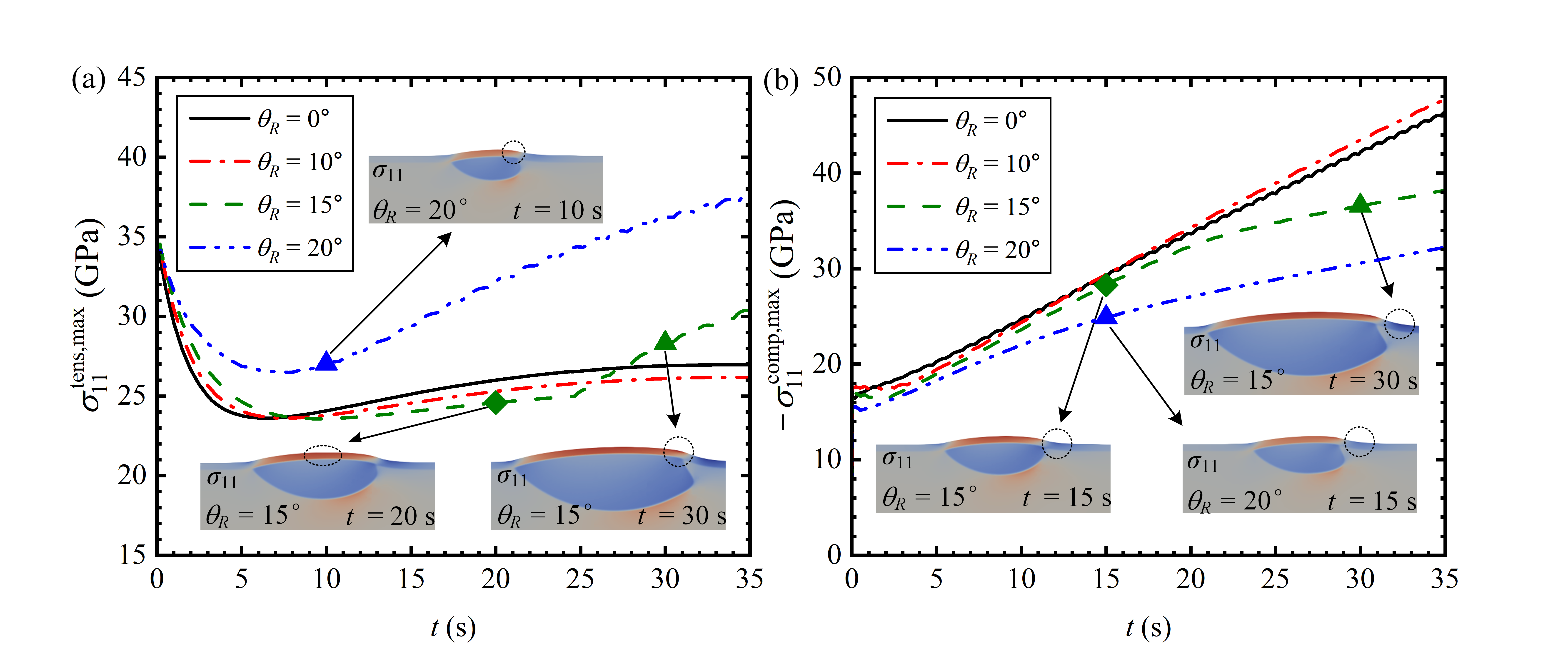}
	\caption{\label{figsingle_orientation_stress}Stress curves during the corrosion processs under different grain orientations $\theta_{R}$: Variation of the maximum (a) tensile  ($\sigma^{\mathrm{tens,max}}_{11}>0$) and (b) compressive ($\sigma^{\mathrm{comp,max}}_{11}<0$) stresses versus time $t$ in the oxide film. Several insets show the stress distribution of triangle and rhombus points on the stress curve, and the dashed circles in the insets show where the maximum stress is located.}
\end{figure*}

\subsubsection{Effect of grain orientation on corrosion morphology} \label{sec3.1.3} 

In studies of hydrides within the zirconium metal, it has been observed that the growth morphology of hydride precipitates varies with grain orientation, as previously reported\cite{han2019phase, heo2019phase}. Grain rotation alters the mechanical symmetry (i.e. eigenstrains and elastic constants) near the metal surface, thereby influencing the corrosion morphology. To accurately assess the corrosion resistance of metallic materials based on their grain microstructure and optimize their performance, it is essential to investigate and understand the impact of grain orientation on the growth of hydride precipitates in hydrogen-induced spot corrosion of metal surfaces. Therefore, we utilize the MPF model to examine the effect of grain orientations on the corrosion morphology during the spot corrosion process of the U monocrystal. To start the simulation, the grain is rotated clockwise by a small angle $\theta_{R}$, and the stress-free strains (Eq. \eqref{eq22}) and elastic constants (Eq. \eqref{eq32}) are adjusted accordingly. The other simulation settings are the same as before.

Figs. \ref{figsingle_orientation_morphology}(a)-(d) illustrate the effect of grain orientation on the morphology outline during the growth of hydride precipitates. Fig. \ref{figsingle_orientation_morphology}(e) shows the variation of the corrosion degree ($V_h/V_0$) of the metal matrix as a function of time $t$ under different grain orientations. Grain rotation exerts an influence on the preferred growth pattern of hydrides\cite{heo2019phase}, inhibiting the lateral growth in the vicinity of the oxide-hydride interface, decreasing the corrosion degree of metal. The effect of grain orientation on the corrosion degree $(V_{h}/V_0)$ can be seen in the Sec. \uppercase\expandafter{\romannumeral2} D of the Supplementary Materials. Figs. \ref{figsingle_orientation_morphology}(e)-(h) show the distribution of the elastic interaction energy $\Delta G^{\mathrm{el}}_{m,h}$ between the metal and hydride at $t=25$ s in the monocrystal for different grain rotation angles. As a result of the rotation, the right-side negative elastic interaction (location 1) is rotated into the lower-right metal, promoting the growth of the hydride into the metal interior, and the positive elastic interaction region (location 2) near the hydride-oxide interface suppresses the hydride growth along the metal-oxide interface.

%%%换成腐蚀形貌？（应力影响见附件）

Furthermore, we investigate the influence of grain orientation on the maximum tensile and compressive stresses within the oxide film, as shown in Fig. \ref{figsingle_orientation_stress}. Several insets show the stress distribution of  triangle and rhombus points on the stress curve, and the dashed circles in the insets show where the maximum stress is located. As the hydride growth progresses, the maximum tensile stress of the oxide film shifts from the center to the vicinity of the slowest point of hydride growth (see the inset when $t=30$ s in Fig. \ref{figsingle_orientation_stress}(a)). For small grain rotation angles ($\theta_{R} = 10^{\circ}$), grain rotation has little effect on the maximum tensile and compressive stresses within the oxide film. However, as the grain rotation angle increases, the influence of grain rotation on these stress parameters gradually becomes significant. For larger angles ($\theta_{R} = 15^{\circ}, 20^{\circ}$), grain rotation further amplifies the maximum tensile stress while reducing the maximum compressive stress. This behavior is directly associated with the variation in hydride morphology induced by grain orientation.

The simulation results reveal that grain rotation induces an asymmetric distribution of elastic interaction energy, which in turn influences the preferred direction of hydride growth and leads to the development of an asymmetric and anisotropic hydride morphology. This morphological change also affects the magnitude and location of the maximum tensile and compressive stresses within the oxide film.

\subsection{Simulation in the uranium bicrystal} \label{sec3.2} 
So far, we have considered {\HC} in the U monocrystal, whereas in the actual metal there are many grains separated by GBs. Experimental evidence suggests that initial hydride nucleation and rapid growth occur at GBs, resulting in the formation of intergranular corrosion\cite{scott2007ud3,banos2018review,banos2016effect,brierley2016anisotropic}. Furthermore, the passage of hydrides through GBs induces interesting morphological transformations\cite{wang2019microstructure}. Therefore, GBs play a critical role in the {\HC} of actual metal materials. In this section, to isolate the interaction of multiple GBs, we investigate the behavior of hydride growth in a U bicrystal containing a pre-existing oxide film and a GB. We explore the infulence of the hydrogen-GB interaction strength ($p$) and the interaction between grain orientation and the GB on the corrosion morphology and stress in the U bicrystal.

\begin{figure*}[htbp]%
	\centering
	\includegraphics[width=0.8\textwidth]{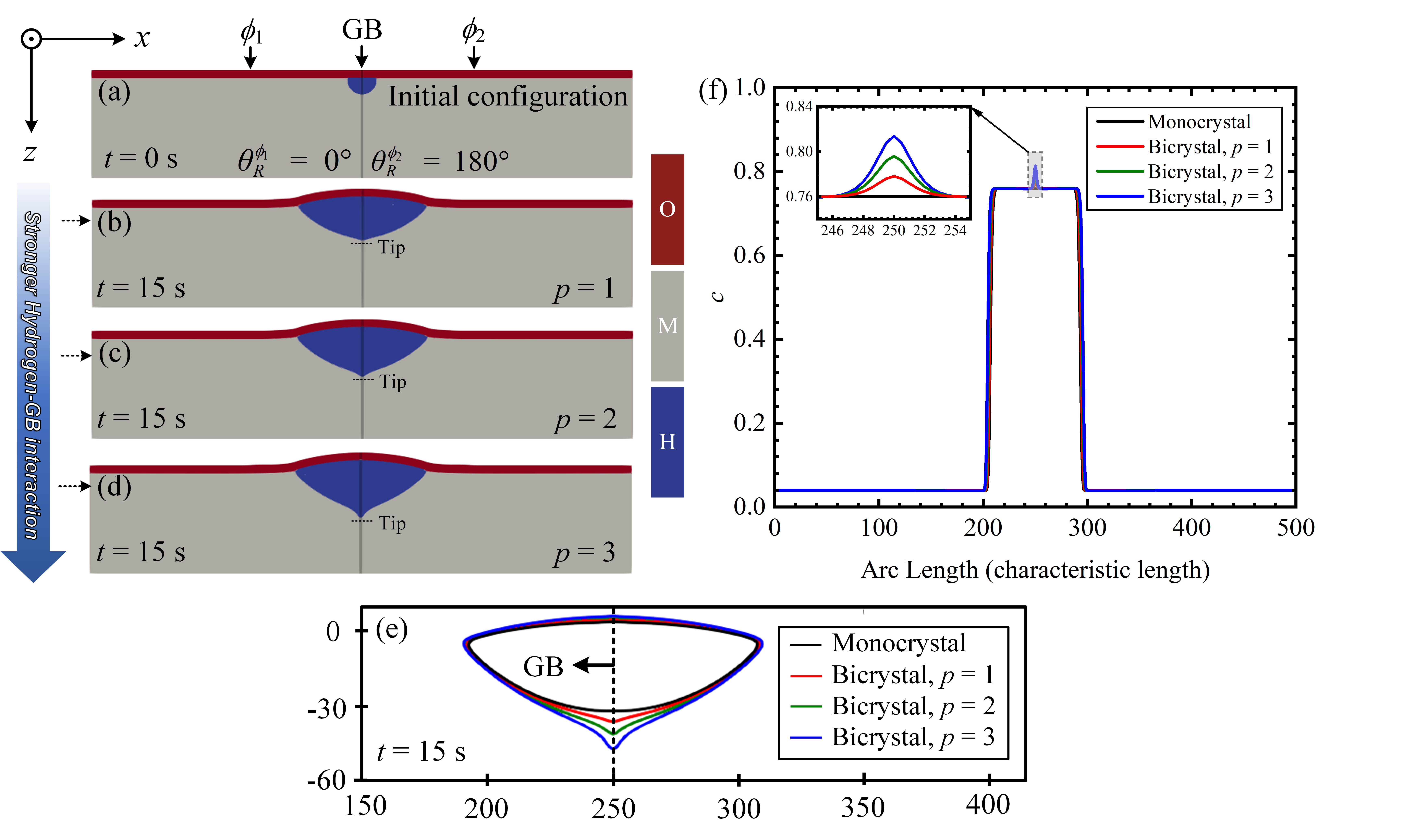}
	\caption{\label{figbicrystal_grain_morphology}Corrosion characteristics under different the hydrogen-GB interaction strength $p$: (a)-(d) Effect of hydrogen-GB interaction on the intergranular corrosion morphology; (e) Effect of hydrogen-GB interaction on the spot corrosion interfaces at $t = 15$ s; (f) Hydrogen concentration distribution along the transverse section marked by the black dot arrow on the left in (b)-(d) at $t = 15$ s.}
\end{figure*}
\begin{figure*}[htbp]%
	\centering
	\includegraphics[width=0.8\textwidth]{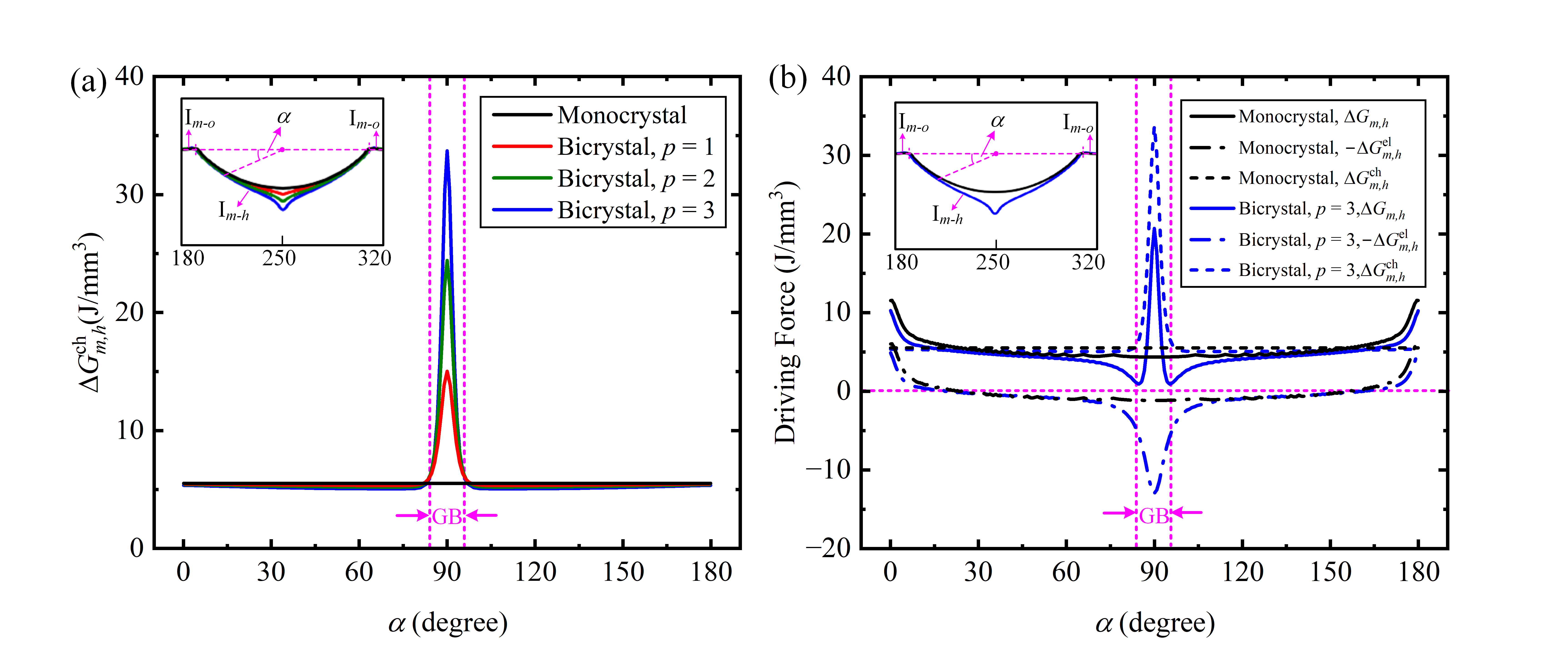}
	\caption{\label{figbicrystal_grain_drivingforce}Driving forces during the corrosion process under different the hydrogen-GB interaction strength $p$: (a) Effects of hydrogen-GB interaction parameters on the chemical driving forces $\Delta G^{\mathrm{ch}}_{m,h}$ at $t=15$ s; (b) Distribution of chemical driving force $\Delta G^{\mathrm{ch}}_{m,h}$, elastic interaction energy $\Delta G^{\mathrm{el}}_{m,h}$ and total driving force $\Delta G_{m,h}$ along the metal-hydride interface (${\rm I}_{m-h}$) at $t=15$s, where ${\rm I}_{m-o}$: metal-oxide interface. }
\end{figure*}

\begin{figure*}[htbp]%
	\centering
	\includegraphics[width=1.05\textwidth]{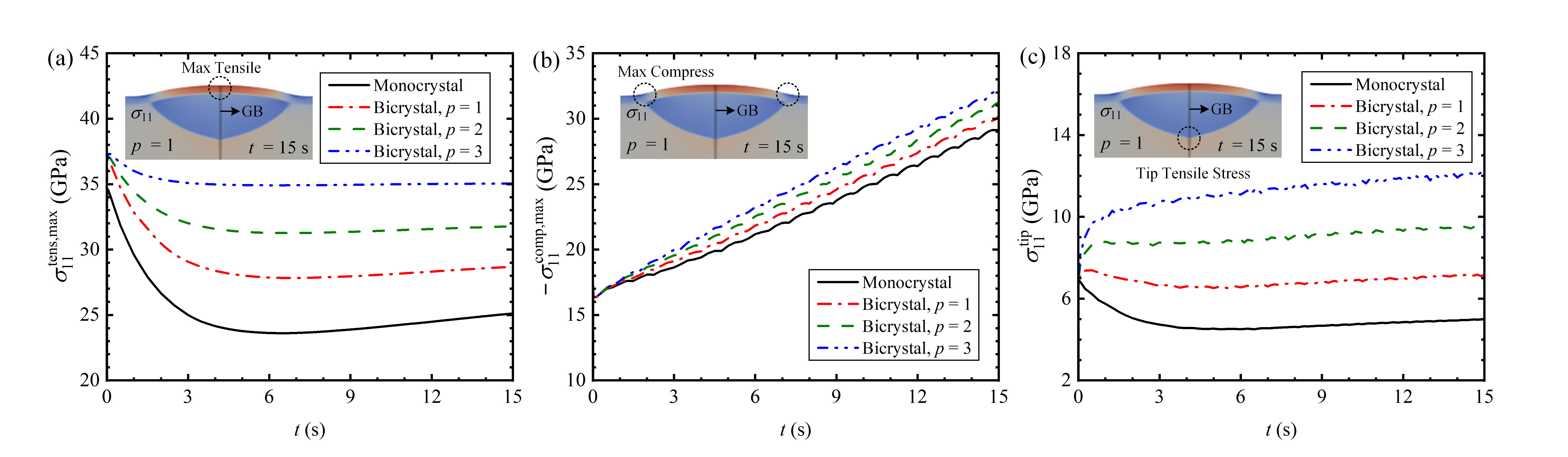}
	\caption{\label{figbicrystal_grain_stress}Stress curves during the corrosion process under different the hydrogen-GB interaction strength $p$: Variation of the maximum (a) tensile  ($\sigma^{\mathrm{tens,max}}_{11}>0$) and (b) compressive ($\sigma^{\mathrm{comp,max}}_{11}<0$) stresses versus time $t$ in the oxide film; (c) Variation of the tip tensile stress versus time $t$ in the metal matrix.}
\end{figure*}

\subsubsection{Effect of GB on intergranular corrosion morphology} \label{sec3.2.1} 
%In the actual metals, the hydrogen concentration tends to be enriched at GBs due to the loose structure there. This phenomenon can be achieved in the PF model by considering the hydrogen-GB interaction\cite{heo2019phase}. The interaction between hydrogen and GBs plays a crucial role in determining the growth behavior of hydrides at GBs, which was overlooked in previous PF models\cite{sheng2022phase,yang2021hydride} of spot corrosion. In our MPF model, we incorporate this hydrogen-GB interaction and investigate its effects on the morphology and stress distribution during corrosion. 
To study GB separately and exclude the interaction between grain orientation and GB, we adopt a method similar to that employed by Heo \textit{et al.}\cite{heo2019phase}. Specifically, the bicrystal system consists of two grains, denoted as $\phi_1$ and $\phi_2$. We set the rotation angles of the two grains in the bicrystal as $\phi_1 = 0^{\circ}$ and $\phi_2 =180^{\circ}$, respectively. This approach allows us to investigate the effects of the GB independently on the intergranular corrosion behavior while minimizing the influence of grain orientation. 

Fig. \ref{figbicrystal_grain_morphology}(a) illustrates the initial conditions of the bicrystal simulation, where a semicircle of initial intergranular hydride precipitation is placed at the GB between adjacent grains. Figs. \ref{figbicrystal_grain_morphology}(b)-(e) show the effect of the hydrogen-GB interaction parameter ($p$) on the intergranular corrosion morphology. The intergranular hydride precipitate in the bicrystal exhibits a distinctive ``tip'' morphology at the GB compared to that observed in the monocrystal. As the hydrogen-GB interaction becomes stronger, this ``tip'' feature becomes more pronounced. Fig. \ref{figbicrystal_grain_morphology}(f) displays the distribution of hydrogen concentration along the transverse section indicated by the dotted arrow in Fig. \ref{figbicrystal_grain_morphology}(b)-(d). The hydrogen concentration is enriched at the GB and exceeds the equilibrium concentration of the hydride inside the grain. In fact, this phenomenon of solute atoms deviating from the equilibrium concentration value inside the GB is called GB segregation. The interaction parameter $p$ is related to the GB segregation potential, which is determined by the GB energy variation with segregating hydrogen at a GB. A stronger hydrogen-GB interaction corresponds to a higher GB segregation potential.

To determine the relationship between the formation of the distinctive ``tip'' morphology at GB and GB segregation, an analysis of the driving forces along the metal-hydride interface is conducted. We define an angle $\alpha$ to describe the distribution of driving forces along the metal-hydride interface. Fig. \ref{figbicrystal_grain_drivingforce}(a) illustrates the effect of the hydrogen-GB interaction parameter on the chemical driving forces at $t=15$ s. The chemical driving force within the GB, induced by the GB segregation, is significantly higher than that inside the grain. Moreover, the chemical driving force within the GB increases with a stronger hydrogen-GB interaction (higher GB segregation potential). Fig. \ref{figbicrystal_grain_drivingforce}(b) shows the distribution of the chemical driving force $\Delta G^{\mathrm{ch}}_{m,h}$, elastic interaction energy $\Delta G^{\mathrm{el}}_{m,h}$, and total driving force $\Delta G_{m,h}$ along the metal-hydride interface at $t=15$ s, comparing the bicrystal and monocrystal scenarios. The total driving force within the GB is higher than that within the grain, primarily due to the higher chemical driving force outweighing the elastic interaction energy, leading to the rapid hydride growth within the GB and the formation of the characteristic ``tip'' morphology.

Figs. \ref{figbicrystal_grain_stress}(a) and (b) depict the effect of the GB on the maximum tensile and compressive stresses within the oxide film, respectively. The introduction of GB increases the maximum tensile and compressive stresses within the oxide film. Moreover, a higher hydrogen-GB interaction parameter increases the maximum tensile stress. Higher interaction parameter values also imply enhanced disorder and mismatch within GBs, leading to increased stress concentration, especially within the GBs of the oxide film. Fig. \ref{figbicrystal_grain_stress}(c) illustrates that the ``tip'' morphology enhances the concentration of tip tensile stress at the bottom of the intergranular hydride precipitate. In addition, experimental evidence suggests that hydride precipitates serve as preferred paths for metal crack growth due to their brittleness\cite{briottet2019industrial,jones2013surface}. Therefore, the concentration of tensile stress near the ``tip'' morphology may play a crucial role in the transition from intergranular spot corrosion to crack corrosion.

Briefly, these simulation results demonstrate that intergranular hydrides located at GBs exhibit a pronounced ``tip'' morphology. Furthermore, the presence of GB increases the maximum tensile and compressive stresses within the oxide film, while the ``tip'' morphology enhances the concentration of tensile stress at the bottom of the hydride precipitate. These results highlight the critical role of GB in the intergranular corrosion behavior.

\begin{figure*}[htbp]%
	\centering
	\includegraphics[width=0.75\textwidth]{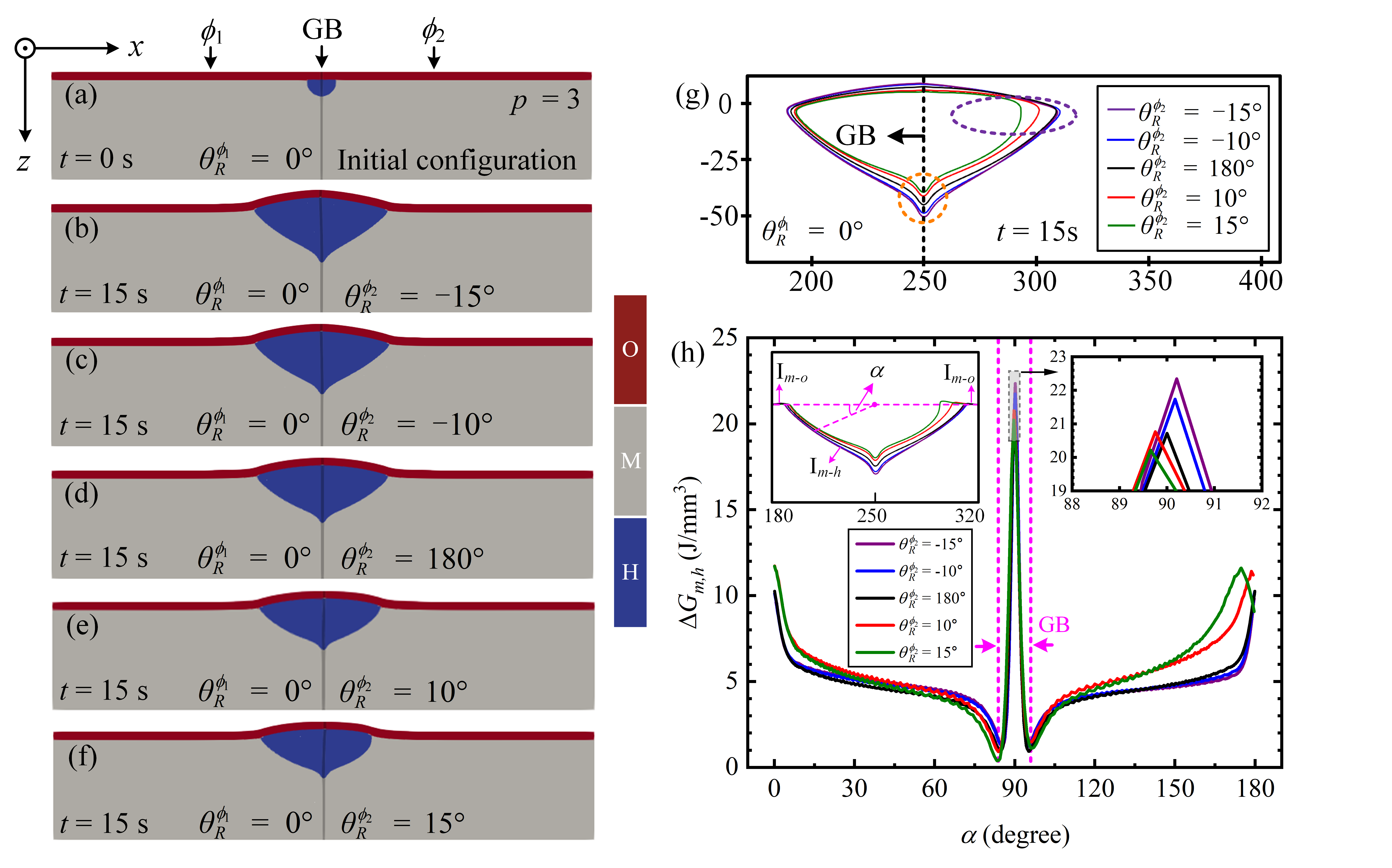}
	\caption{\label{figbicrystal_grain_orientation_morphology}Corrosion characteristics under the interaction between different grain orientations $\theta^{\phi{2}}_{R}$ and GB: (a)-(e) Effect of the interaction between different grain orientations and GB on the intergranular corrosion morphology. (f) Outline of hydride precipitates in (b)-(e) (g) Distribution of total driving force $\Delta G_{m,h}$ along the metal-hydride interface (${\rm I}_{m-h}$) under different grain orientations at $t=15$ s, where ${\rm I}_{m-o}$: metal-oxide interface. }
\end{figure*}

\begin{figure*}[htbp]%
	\centering
	\includegraphics[width=1.05\textwidth]{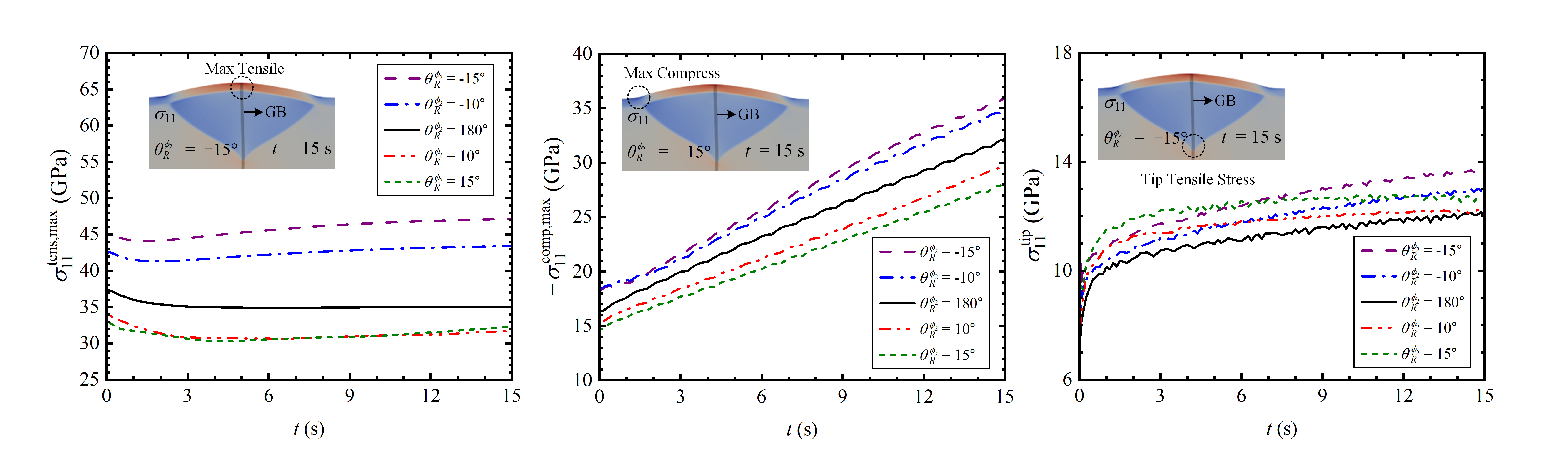}
	\caption{\label{figbicrystal_grain_orientation_stress}Stress curves during the corrosion process under the interaction between different grain orientations $\theta^{\phi{2}}_{R}$: Variation of the maximum (a) tensile  ($\sigma^{\mathrm{tens,max}}_{11}>0$) and (b) compressive ($\sigma^{\mathrm{comp,max}}_{11}<0$) stresses versus time $t$ in the oxide film; (c) Variation of the tip tensile stress versus time $t$ in the metal matrix.}
\end{figure*}

\subsubsection{Effect of the interaction between grain orientation and GB on intergranular corrosion morphology} \label{sec3.2.3} 

In the following simulation, we extended the investigation by adjusting the grain orientation to simulate the influence of the interaction between different grain orientations and GB on the intergranular hydride growth. In the bicrystal system, the rotation angle of grain $\phi{1}$ remains constant, grain $\phi_2$ is rotated clockwise by different angles ($\theta^{\phi{2}}_{R} = -15^{\circ}, -10^{\circ}, 180^{\circ}, 10^{\circ}, 15^{\circ}$). This approach allows for an examination of the effect of the interaction between grain orientation and GB on the intergranular corrosion behavior of the system.

Figs. \ref{figbicrystal_grain_orientation_morphology}(a)-(f) illustrate the effect of the interaction between different grain orientations and GB on the intergranular corrosion morphology. The corresponding hydride precipitate outlines for images (b)-(f) are displayed in Fig. \ref{figbicrystal_grain_orientation_morphology}(g). Two important features can be observed from the results: 1) The grain rotation changes the symmetry of the intergranular hydride morphology similar to the moncrystal case, as indicated by the purple dashed ellipse in Fig. \ref{figbicrystal_grain_orientation_morphology}(g); 2) The growth rate of intergranular hydrides inside the GB is accelerated for counterclockwise grain rotation ($\theta^{\phi{2}}_{R}<0^{\circ}$) and decelerated for clockwise grain rotation ($\theta^{\phi{2}}_{R}>0^{\circ}$), as shown by the orange dashed circle in Fig. \ref{figbicrystal_grain_orientation_morphology}(f).

The formation of the two significant features mentioned above is attributed to the modification of the distribution of total driving forces along the metal-hydride interface induced by the grain orientation and its interaction with GB. We examine the distribution of the total driving force along the metal-hydride interface at $t=15$ s, as shown in Fig. \ref{figbicrystal_grain_orientation_morphology}(h). The angle $\alpha$ is adopted as a variable to describe the distribution of driving forces along the interface. The rotation of grain $\phi_2$ leads to a distinct difference in the distribution of the total driving force along the metal-hydride interface within this grain compared to the adjacent grain $\phi_1$. Consequently, an overall asymmetric distribution of the total driving force is formed along the metal-hydride interface,  thereby resulting in the intergranular hydride exhibiting pronounced asymmetry. The asymmetry is particularly pronounced for the clockwise rotated grains (see Fig. \ref{figbicrystal_grain_orientation_morphology}(g)). Moreover, the interaction between grain orientation and GB adjusts the magnitude of the total driving force within the GB, as observed in the magnified view of Fig. \ref{figbicrystal_grain_orientation_morphology}(h). Counterclockwise grain rotation ($\theta^{\phi{2}}_{R} < 0^{\circ}$) leads to an increase in the total driving force at $\alpha = 90^\circ$, amplifying the prominence of tip morphology. Conversely, when the grain rotates clockwise ($\theta^{\phi{2}}_{R} > 0^{\circ}$) at $\alpha = 90^\circ$, the total driving force decreases, resulting in a slower growth rate of the tip morphology. These findings emphasize the significance of grain orientation and its interaction with GB in modifying the morphology of intergranular hydrides.

Fig. \ref{figbicrystal_grain_orientation_stress} shows the influence of the interaction between different grain orientations and GB on the stresses within the oxide film. This interaction leads to an increase in both the maximum tensile and compressive stresses for counterclockwise grain rotation ($\theta^{\phi{2}}_{R}<0^{\circ}$), while it decreases these stresses for clockwise grain rotation ($\theta^{\phi{2}}_{R}>0^{\circ}$). Additionally, the interaction results in an increase in the tip tensile stress within the metal. The increase in tip tensile stress caused by clockwise grain rotation is less than that caused by counterclockwise grain rotation.

From the results, the grain orientation and its interaction with the GB not only lead to the formation of asymmetric intergranular hydride morphology, but also affect the growth rate of hydrides inside the GB and the stresses in the oxide film. Counterclockwise grain rotation accelerates hydride growth at the GB and increases the tensile and compressive stresses, while clockwise grain rotation decelerates hydride growth and decreases these stresses. These observations highlight the important influence of the mismatch direction and angle between the grain orientations on the intergranular corrosion morphology and stress distribution, and provide valuable insights into the understanding of intergranular corrosion mechanisms.

%These observations highlight the importance of intricate interplay between grain rotation and GB on the corrosion morphology and stress distribution. 

\begin{table}[ht]%table2
	\caption{\label{tab2} Grain rotation angle $\theta^{\phi_{g}}_{R}$ used in the uranium polycrystal simulations.}
	\begin{tabular}{lc}
		\hline
		Grain  $\qquad\qquad$ & Rotation Angle $\theta^{\phi_{g}}_{R}$        \\
		\hline
		gr0     & $0^{\circ}$                        \\                                
		gr1     & $10^{\circ}$                       \\
		gr2     & $-10^{\circ}$                      \\
		gr3     & $15^{\circ}$                       \\
		gr4     & $-15^{\circ}$                      \\
		\hline
	\end{tabular}	
\end{table}

\begin{figure*}[htbp]%
	\centering
	\includegraphics[width=0.8\textwidth]{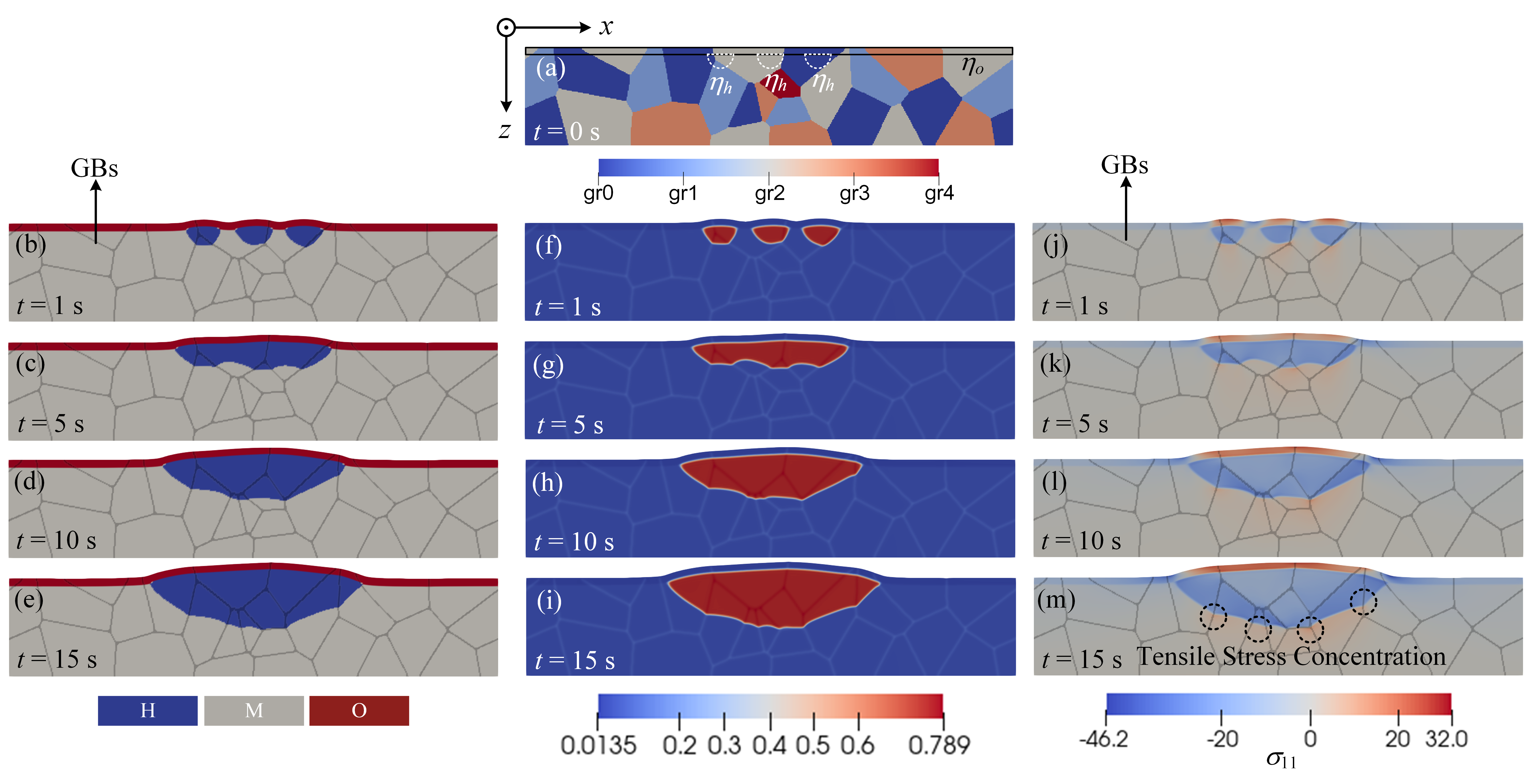}
	\caption{\label{fig_multi_spots_polycrystal_morphology}Corrosion characteristics during the multi-spot corrosion process in the polycrystal: (a) Grain structure diagram of the polycrystal for the simulation, and the positions of the corrosion spots and oxide film are marked. (b)-(e) Evolution of multi-spot corrosion morphology. (f)-(i) Variation of distribution of hydrogen concentration during the corrosion process versus time $t$. (j)-(m) Variation of distribution of stress $\sigma_{11}$ during the corrosion process versus time $t$.}
\end{figure*}

\begin{figure*}[htbp]%
	\centering
	\includegraphics[width=0.8\textwidth]{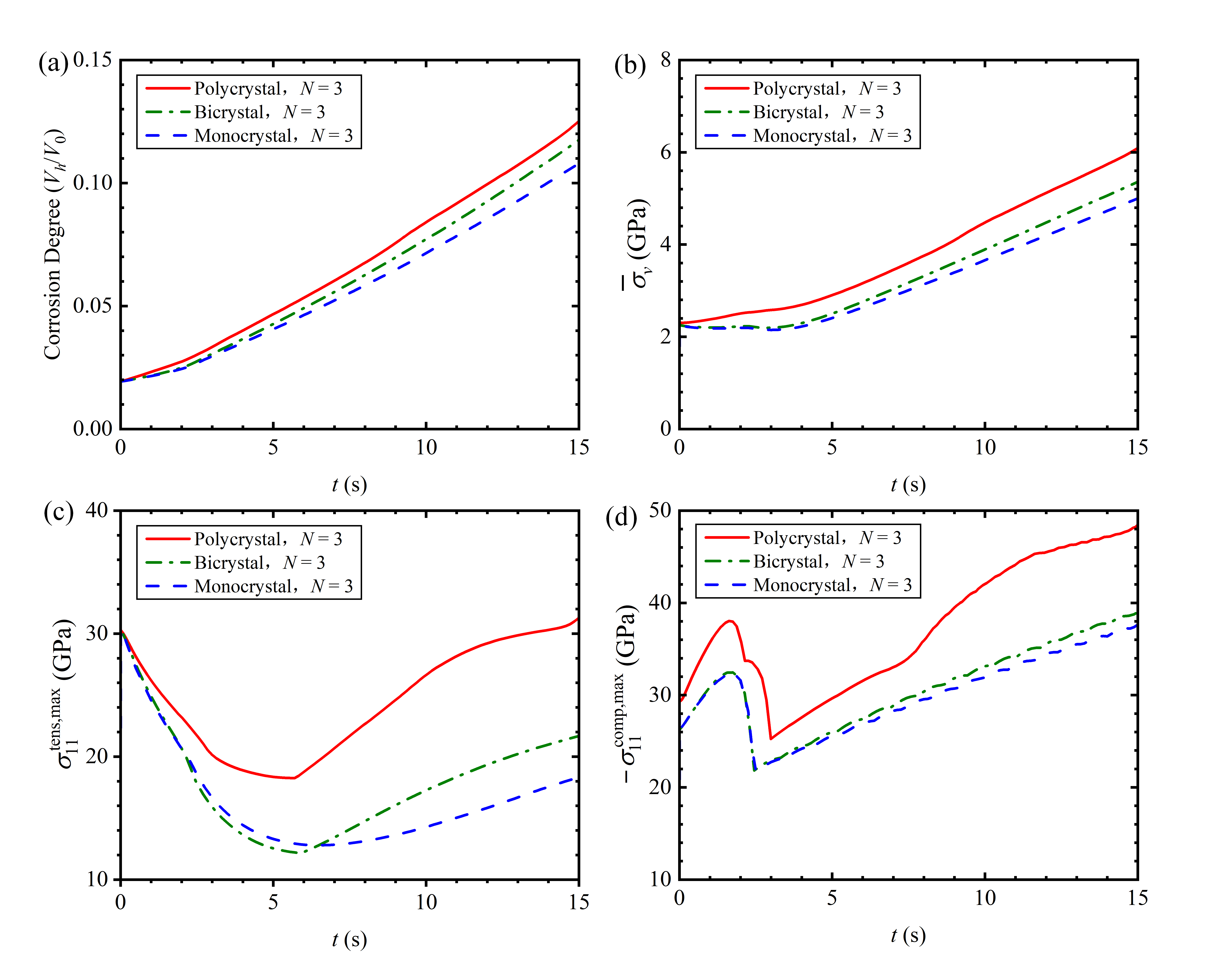}
	\caption{\label{figpolycrystal_morphology_statistics}Corrosion and stress curves during the multi-spot corrosion process in the polycrystal: (a) Variation of the corrosion degree $(V_{h}/V_0)$ of the metal matrix versus time $t$; (b) Variation of the average von Mises stress $\bar{\sigma}_v$ of the materials versus time $t$; Variation of the maximum (c) tensile  ($\sigma^{\mathrm{tens,max}}_{11}>0$) and (d) compressive ($\sigma^{\mathrm{comp,max}}_{11}<0$) stresses versus time $t$ in the oxide film.} 
\end{figure*}

\subsection{Simulation in the uranium polycrystal} \label{sec3.3}
Finally, we perform more complex polycrystalline simulations that are closer to the real situation, and compare the corrosion degree and the material's internal stresses in polycrystals with the bicrystal and monocrystal counterparts. For simplicity, the grain orientation of the compared monocrystal is $\theta_{R} = 0^{\circ}$, and the grain orientations of the compared bicrystal are $\theta^{\phi{1}}_{R} = 0^{\circ}$ and $\theta^{\phi{2}}_{R} = 180^{\circ}$. The grain structure diagram shown in Fig. \ref{fig_multi_spots_polycrystal_morphology}(a) is employed for the polycrystalline simulation, with the rotation angles $\theta^{\phi_{g}}_{R}$ specified for each grain in Table \ref{tab2}. At the beginning of the simulation, three identical semicircular initial hydride precipitates are positioned at the top of the U polycrystal. Note that the average grain size of the grain structure in our computational domain is smaller than that of realistic polycrystalline U metals. However, we also note that the generated grain structure is sufficient for the purpose of conducting demonstrative simulations as it encompasses all the key physical factors discussed above.

The evolution of the multi-spot corrosion morphology within the U polycrystal is demonstrated in Figs. \ref{fig_multi_spots_polycrystal_morphology}(b)-(e). The increased number of GBs and grain orientations contribute to the emergence of more ``tip'' and irregular morphology, which is close to the real corrosion morphology in the experiments\cite{brierley2016anisotropic,ji2019mechanism}. These ``tip'' morphologies emerge as a result of the enhanced chemical driving force induced by GB segregation at the GBs, as illustrated in Figs. \ref{fig_multi_spots_polycrystal_morphology}(f)-(i). In addition, Figs. \ref{fig_multi_spots_polycrystal_morphology}(j)-(m) display the monitoring of internal stress $\sigma_{11}$ during multi-spot corrosion. The presence of more ``tips'' creates more tensile stress concentrations in the metal matrix, while the irregular bulge of the oxide film results in multiple tensile stress concentration locations within the oxide film. These stresses may play a critical role in the degradation and eventual fracture of the oxide film at multiple local locations, ultimately culminating in the overall exfoliation of the oxide film. 

%The von Mises stress, defined as $\sigma_v = \left\{(1/2)\left[(\sigma_{11}-\sigma_{22})^2+(\sigma_{22}-\sigma_{33})^2+(\sigma_{33}-\sigma_{11})^2\right.\right.$  \\  $\left.\left. +6(\sigma_{12}^2+\sigma_{23}^2+\sigma_{31}^2)\right]\right\}^{1/2}.$, is an important parameter in assessing material yield and fracture in engineering. 

The variation of the corrosion degree and the volume average von Mises stress of the system as a function of time are illustrated in Figs. \ref{figpolycrystal_morphology_statistics}(a) and (b), respectively. The calculated corrosion degree and the volume average von Mises stress in the polycrystalline system are higher than those in the bicrystal and monocrystal cases. This disparity can be attributed to the presence of a large number of GBs and diverse grain orientations in the polycrystal, which accelerate the corrosion process. Figs. \ref{figpolycrystal_morphology_statistics}(c) and (d) provide detailed information on the maximum tensile and compressive stresses in the oxide film as a function of time, respectively. The oxide film in the polycrystal exhibits higher stresses compared to the monocrystal and bicrystal cases, which may increase the fracture susceptibility of the oxide film in the polycrystal. This phenomenon is attributed to the irregular bulge of the oxide film induced by the multiple grain orientations. The irregular bulge increase the stress level within the oxide film, resulting in intensified stress concentration near the GBs within the oxide film. These findings highlight the role of complex microstructures in determining the stress evolution during corrosion and the corrosion behavior in polycrystalline materials.

%This phenomenon can be attributed to the accelerated growth of hydrides, leading to enhanced stress concentration near the GBs inside oxide film.

Our simulations successfully reproduced intricate and irregular spot morphologies and captured the dynamic stress changes within the materials. These results demonstrate the comprehensiveness and effectiveness of our MPF model in addressing complex corrosion phenomena.

\section{Conclusion}\label{sec4} 
In conclusion, we develop a comprehensive MPF model incorporating i) hydrogen diffusion, ii) phase transformations, iii) elastic interactions between phases, iv) elastic interactions between grains, and v) interactions between hydrogen solutes and GBs for simulating {\HC} in the rare metal surface. This MPF model can address the limitations of conventional PF models by introducing the oxide film phase, resulting in more accurate and realistic hydride morphologies with sharp edges, which aligns well with experimental observations. Moreover, this MPF model also includes multiple grains and GBs, allowing for the simulation of multi-spot corrosion in polycrystals with complex textures to obtain an irregular corrosion morphology that is closer to the real situation. Our MPF model provides a valuable tool for studying the complex {\HC} behavior, contributing to a better understanding of the corrosion process in practical materials.

We apply the model to the study of {\HC} in the U metal surface with a pre-existing oxide film, and systematically investigate the microscopic mechanisms affecting the spot corrosion morphology and the internal stress state of materials. In the monocrystalline system, we investigate the effects of oxide film thickness, merging of corrosion spots, and grain orientation on the corrosion morphology and stress.

1) It was observed that a thick oxide film slows down the hydride growth, reduces the tensile stress in the center of the oxide film, and reduces compressive stress in the film near the edge of the hydride. This is attributed to the lower elastic energy and higher deformation resistance of the thick oxide film.

2) Negative elastic interaction between adjacent corrosion spots leads to the merging of corrosion spots, which reduces the maximum tensile stress in the oxide film and releases compressive stress between adjacent spots. 

3) Grain rotation influences the preferred growth direction of hydrides, with lateral growth inhibited near the oxide-hydride interface, resulting in a shift of the maximum stress point from the center of the oxide film to the inhibition point. This is attributed to the effect of grain rotation on the elastic interaction energy distribution.

In the bicrystal study, we investigated the effects of GB and multiple grain orientations on the intergranular corrosion morphology and internal stresses.

1) We observed the enrichment of the hydrogen concentration at the GB, producing the GB segregation. As a result, the intergranular hydride grows rapidly along the GB, forming a ``tip'' morphology at the GB. 

2) Grain orientation and its interaction with the GB induce the formation of asymmetric intergranular hydride and influence the hydride growth rate insdie the GB. In addition, this interaction increases the tensile and compressive stresses in the oxide film for counterclockwise grain rotation, while slowing the growth rate and decreasing these stresses for clockwise grain rotation.

Finally, in the polycrystalline study, we consider the comprehensive effects of these phenomena and factors. The results demonstrate that the complex and irregular morphology in polycrystalline materials leads to higher stress levels and corrosion degree compared to the monocrystal and bicrystal cases. This is attributed to the presence of a large number of GBs and diverse grain orientations in polycrystals. These findings emphasize the role of complex microstructures in determining stress and corrosion behavior in polycrystalline materials. 

These obtained results serve to validate the efficacy and accuracy of the MPF model within the {\HC}, offering potential applications in corrosion mitigation and control strategies. However, the current MPF model primarily focuses on the corrosion morphology and stress distribution in the early growth stage of {\HC}, without explicitly considering the plastic strain of metal and fracture process of the oxide film during the corrosion process. To fully capture the metal plasticity and fracture behavior of the oxide film for extending the model to the oxide cracking stage of {\HC}, the metal plastic hardening and dynamic fracture mechanisms of the oxide film can be considered into the MPF model in future studies.

\section{Acknowledgement}
We thank Yuan-Ji Xu, Xing-Yu Gao, Dan Jian, Li-Fang Wang and Le Zhang for the help with basic knowledge and properties of uranium. We thank Chao Yang and Guo-Min Han for the help in the method of modeling. We thank Bei-Lei Liu, Kai-Le Chen, Ji-De Zou, Yu Song for helpful discussions. The work is supported by the National Key R\&D Program of China (No. 2021YFB3501503), the National Natural Science Foundation of China (No. 12004048), the Science Challenge Project (No. TZ2018002) and the Foundation of LCP. We thank the Tianhe platforms at the National Supercomputer Center in Tianjin.

\section{Author Contributions}
H.-F. Song and Y. Liu conceived and supervised the project. J. Sheng performed the numerical simulations. All authors analyzed and discussed the results. J. Sheng, Y. Liu, X.-M. Shi, Y.-C. Wang and H.-F. Song wrote the manuscript, with contributions from all the authors.

\section{Data availability}
The data that support the findings of this study are available from the corresponding author upon reasonable request. The copyright of the experimental images has been applied for on the corresponding journal website.

\bibliographystyle{elsarticle-num}	
\bibliography{main_PF2_and_SI}

\clearpage

\end{document}

\title{Supplementary Materials to ``A multiphase-field model for simulating the hydrogen-induced multi-spot corrosion on the surface of polycrystalline metals: Application to uranium metal''}

\author{Jie Sheng}
\affiliation{Laboratory of Computational Physics, Institute of Applied Physics and Computational Mathematics, Beijing 100088, China}

\author{Yu Liu}
\email{liu\_yu@iapcm.ac.cn}
\affiliation{Laboratory of Computational Physics, Institute of Applied Physics and Computational Mathematics, Beijing 100088, China}

\author{Xiao-Ming Shi}
\affiliation{School of Materials Science and Engineering,Beijing Institute of Technology, Beijing 100081, China}
\affiliation{Advanced Research Institute of Multidisciplinary Science, Beijing Institute of Technology, Beijing 100081, China}

\author{Yue-Chao Wang}
\affiliation{Laboratory of Computational Physics, Institute of Applied Physics and Computational Mathematics, Beijing 100088, China}

\author{Zi-Hang Chen}
\affiliation{Laboratory of Computational Physics, Institute of Applied Physics and Computational Mathematics, Beijing 100088, China}
\affiliation{School of Materials Science and Engineering,Beijing Institute of Technology, Beijing 100081, China}
\affiliation{Advanced Research Institute of Multidisciplinary Science, Beijing Institute of Technology, Beijing 100081, China}

\author{Ke Xu}
\affiliation{Laboratory of Computational Physics, Institute of Applied Physics and Computational Mathematics, Beijing 100088, China}
\affiliation{School of Materials Science and Engineering,Beijing Institute of Technology, Beijing 100081, China}
\affiliation{Advanced Research Institute of Multidisciplinary Science, Beijing Institute of Technology, Beijing 100081, China}

\author{Shuai Wu}
\affiliation{Laboratory of Computational Physics, Institute of Applied Physics and Computational Mathematics, Beijing 100088, China}
\affiliation{School of Materials Science and Engineering,Beijing Institute of Technology, Beijing 100081, China}
\affiliation{Advanced Research Institute of Multidisciplinary Science, Beijing Institute of Technology, Beijing 100081, China}

\author{Hou-Bing Huang}
\affiliation{School of Materials Science and Engineering,Beijing Institute of Technology, Beijing 100081, China}
\affiliation{Advanced Research Institute of Multidisciplinary Science, Beijing Institute of Technology, Beijing 100081, China}

\author{Bo Sun}
\affiliation{Laboratory of Computational Physics, Institute of Applied Physics and Computational Mathematics, Beijing 100088, China}

\author{Hai-Feng Liu}
\affiliation{Laboratory of Computational Physics, Institute of Applied Physics and Computational Mathematics, Beijing 100088, China}

\author{Hai-Feng Song}
\email{song\_haifeng@iapcm.ac.cn}
\affiliation{Laboratory of Computational Physics, Institute of Applied Physics and Computational Mathematics, Beijing 100088, China}

\pacs{81.65.Kn, 05.70.Np, 81.40.Np}
\date{\today}
\maketitle

%This supplementary materials show the some calculation details of different phases of free energy density ($f_m(c_m)$, $f_h(c_h)$, $f_o(c_o)$), and some other interesting simulation results. The supplementary materials are organized as follows: Sec. \ref{sec_parameter} shows how to determine $A_{m}$, $A_{h}$, $A_{o}$ and $B_{o}$ by constructing the common tangent. Sec. \ref{sec_elastic_energy_density} shows the effect of initial shape of corrosion spot on the average elastic energy density of oxide film. Sec. \ref{sec_distance} illustrates the effect of the distance between the corrosion spots on the corrosoin morphology. Sec. \ref{sec_theta_GB} demonstrates the influence of angle between the grain boundary and metal surface on the corrosion morphology. Sec. \ref{sec_comparsion} shows the qualitative comparison between phase-field simulations and experimentally observed hydride microstructures. Sec. \ref{sec_stress_monocrystal} shows the effect of thickness of oxide film, multi-spot corrosion and grain orientation on stresses $\sigma_{22}$ and $\sigma_{12}$ in the monocrystal simulation. Sec. \ref{sec_stress_bicrystal} shows the effect of grain boundary, the angle between grain boundary and surface, and the interaction between grain orientation and grain boundary on stresses $\sigma_{22}$ and $\sigma_{12}$ in the bicrystal simulation.

\section{Determining of $A_m$, $A_h$, $A_o$ and $B_o$}\label{sec_parameter}

For determining the parameters $A_m$, $A_h$, $A_o$ and $B_o$, the common tangent line between different phases is constructed. A total of two ($m-h$ and $m-o$) common tangent lines are needed. For the $m-h$ common tangent line, the equilibrium H content of the common tangent point between metal matrix and hydride precipitate are $c_{m}^{\mathrm{mh,eq}} =  2.32$ at.\% and $c_{h}^{\mathrm{mh,eq}} = 74.99$ at.\%\cite{morrell2013uranium}. To the m-o common tangent line, it is assumed that the equilibrium H content of the common tangent point between metal matrix and oxide film are $c_{m}^{\mathrm{mo,eq}} = 5$ at.\% and $c_{o}^{\mathrm{mo,eq}} = 3.12$ at.\%. The calculated method of equilibrium H content $c_{m}^{\mathrm{mh,eq}}$ and $c_{h}^{\mathrm{mh,eq}}$ can be find in the Morrell's work\cite{morrell2013uranium}.

According to the property of the common tangent line, we can obtain two linear equations about $A_m$ and $A_h$ below:
\begin{align}
 	\left. {\partial f_{m} \over \partial c_m}\right|_{c_m = c_m^{\mathrm{mh,eq}}} =   \left. {\partial f_{h} \over \partial c_h}\right|_{c_h = c_h^{\mathrm{mh,eq}}},\\
 	\left. {\partial f_{m} \over \partial c_m}\right|_{c_m = c_\alpha^{\mathrm{mh,eq}}} ={f_{m}(c_\alpha^{\mathrm{mh,eq}}) - f_{h}(c_h^{\mathrm{mh,eq}})\over c_m^{\mathrm{mh,eq}}-c_h^{\mathrm{mh,eq}}}.
\end{align}
Therefore, $A_m$ and $A_h$ can be determined by solving the equations above.

Similarly, we have other two linear equations about $A_o$ and $B_{o}$ below:
\begin{align}
	\left. {\partial f_{m} \over \partial c_m}\right|_{c_m = c_m^{\mathrm{mo,eq}}} =   \left. {\partial f_{o} \over \partial c_o}\right|_{c_o = c_o^{\mathrm{mo,eq}}},\\
	\left. {\partial f_{m} \over \partial c_m}\right|_{c_m = c_\alpha^{\mathrm{mo,eq}}} ={f_{m}(c_\alpha^{\mathrm{mo,eq}}) - f_{o}(c_o^{\mathrm{mo,eq}})\over c_m^{\mathrm{mo,eq}}-c_o^{\mathrm{mo,eq}}}.
\end{align}
Therefore,  $A_o$ and $B_{o}$ can be determined by solving the equations above.

\section{Some additional researches on the Monocrystal}

\subsection{Evolution of spot corrosion morphology under different oxide film thicknesses}\label{sec_evolution_filmthickness}

\begin{figure*}[htbp]%
	\centering
	\includegraphics[width=1.0\textwidth]{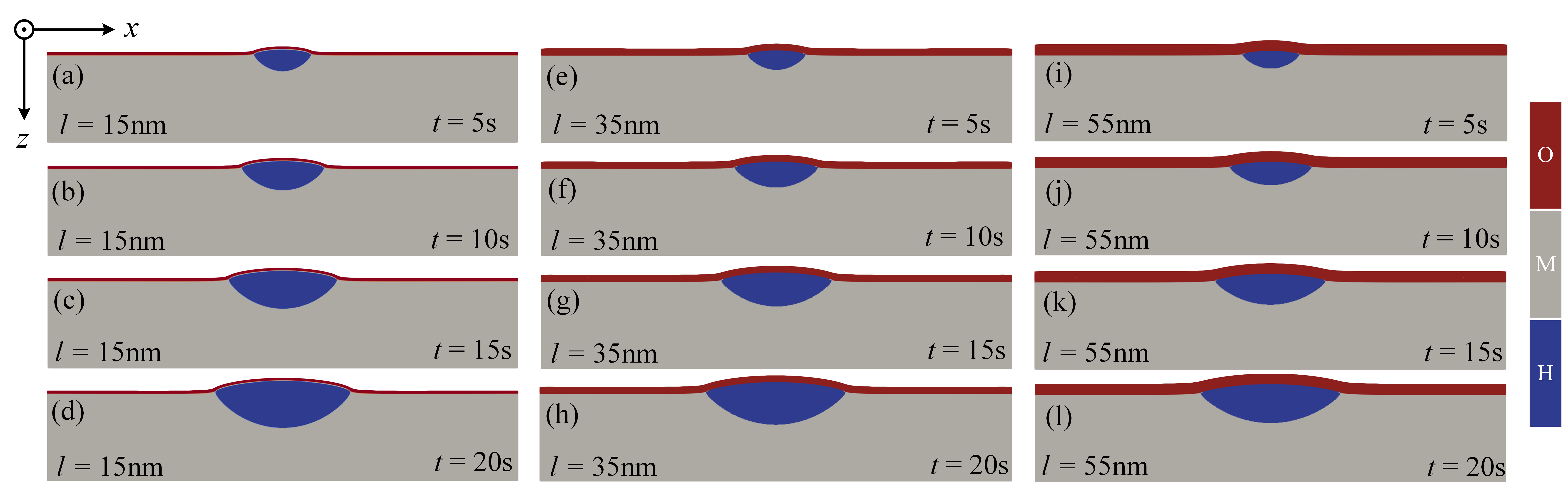}
	\caption{\label{figevolution_spots_filmthickness}Evolution of spot corrosion morphology under different oxide film thicknesses}
\end{figure*}

Fig. \ref{figevolution_spots_filmthickness} demonstrates the evolution of spot corrosion morphology under different oxide film thicknesses. Hydride precipitate exhibits an anisotropic growth behavior, forming a flat shape. As the hydride precipitates expand and grow, the oxide film phase gradually bulges. In additon, an increased thickness of the oxide film cause a reduction in the size of the hydride precipitates, as well as a diminished magnitude of the surface bulge.

\begin{figure*}[htbp]%
	\centering
	\includegraphics[width=0.5\textwidth]{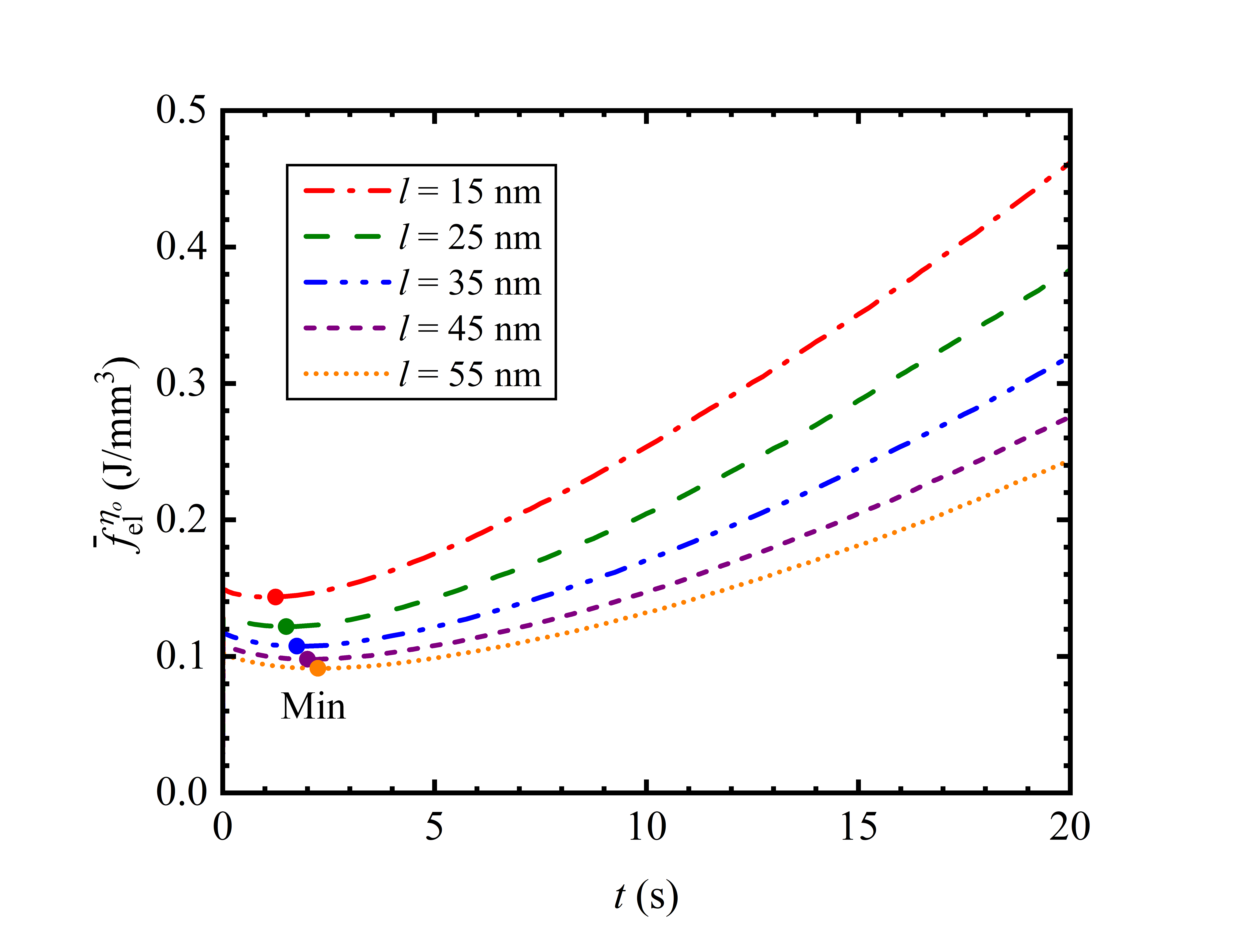}
	\caption{\label{figaverage_elastic_strain_energy_under_different_oxide_film_thicknesses}Effect of the oxide film thickness on the average elastic energy density of oxide film}
\end{figure*}

\subsection{Morphological conversion of hydride causing a decrease in the average elastic energy density of the oxide film}\label{sec_elastic_energy_density}

Fig. \ref{figaverage_elastic_strain_energy_under_different_oxide_film_thicknesses} depicts the variation of the average elastic energy density $\bar{f}^{\eta_o}_{\rm el}$ of the oxide film phase as a function of time $t$ under different oxide film thicknesses $l$. The average elastic energy density initially decreases with time, which corresponds to the process of the morphological conversion of the hydride from a non-thermodynamic stable semicircle to an thermodynamic stable ellipse under the elastic interaction between phases. Subsequently, the continuous increase in the average elastic energy density is attributed to the growth of the hydride. In addition, the average elastic energy diminishes as the thickness of the oxide film increases. This can be attributed to the reduction in local strain and stress at material points within the oxide film and the reduction of deformation. In short, a lower average elastic energy stored within the oxide film signifies a smaller volume of hydride located beneath it. These finding highlights the role of the oxide film as a protective barrier against corrosion.

\begin{figure*}[htbp]%
	\centering
	\includegraphics[width=1.0\textwidth]{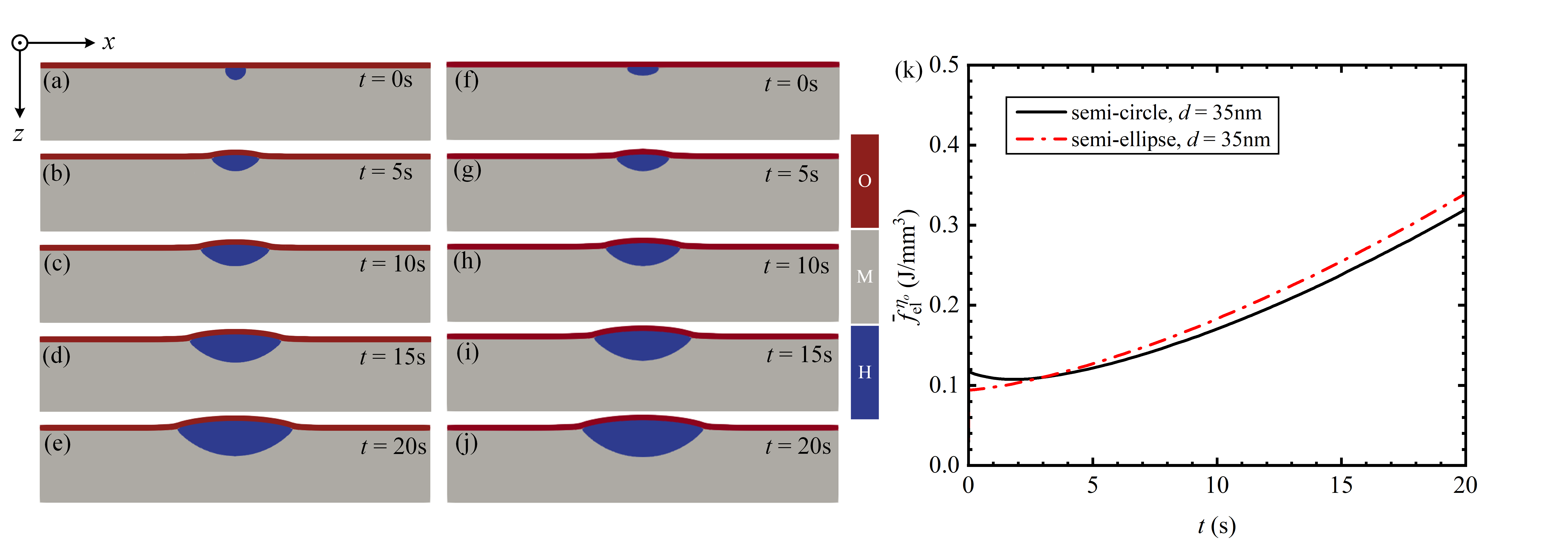}
	\caption{\label{figaverage_elastic_strain_energy}Effect of initial shape of corrosion spot on the evolution of corrosion morphology and average elastic energy density of oxide film}
\end{figure*}

To verify that the initial decrease in the average elastic energy density  $\bar{f}^{\eta_o}_{\rm el}$ is the result of the morphological conversion of the hydride, the influence of the initial shape of corrosion spots on corrosion morphology and the average elastic energy density  $\bar{f}^{\eta_o}_{\rm el}$ of the oxide film is examined, as depicted in Fig. \ref{figaverage_elastic_strain_energy}. It can be observed that the average elastic energy density of the oxide film  in the system with a semi-elliptical hydride as the initial nucleation shape does not exhibit the decreasing trend. This indirectly proves that the morphological conversion of hydride indeed affect the decrease of the average elastic energy density of the oxide film. Furthermore, we also observe that the initial nucleation shape has minimal impact on the final corrosion morphology, which suggests that the final thermodynamic steady state of the system does not depend on the shape of the initial nucleation, but only on the properties of the material itself. This further proves that our simulation results are real and valid and conform to the laws of physics.

\subsection{Effect of the distance between and the number of the corrosion spots on the corrosion morphology and stresses $\sigma_{22}$ and $\sigma_{12}$}\label{sec_distance}

\begin{figure*}[htbp]%
	\centering
	\includegraphics[width=1.0\textwidth]{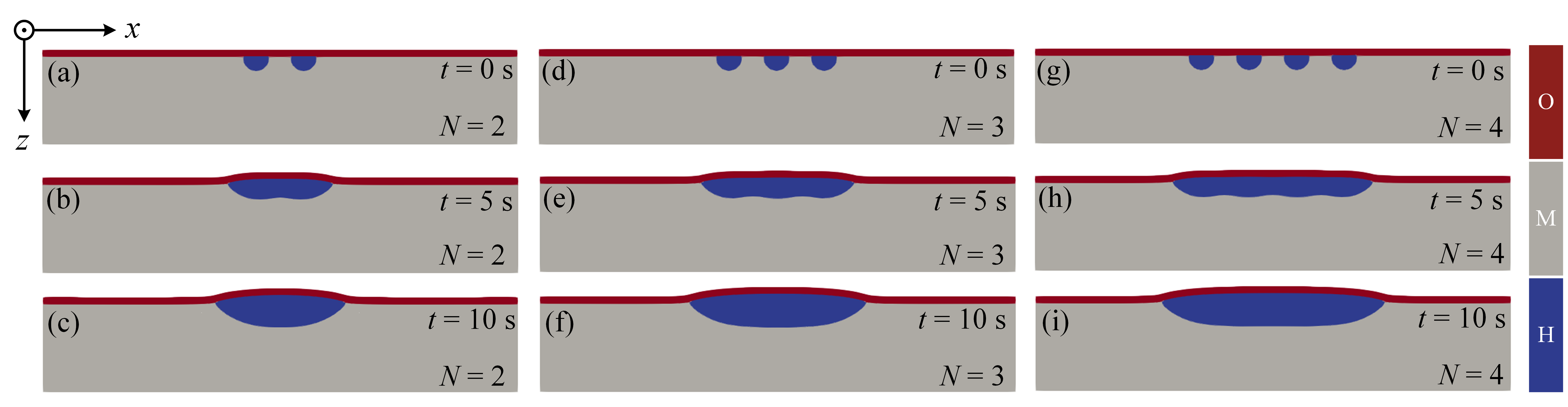}
	\caption{\label{figmultiple_spots_number}Effect of the number of corrosion spots on the evolution of corrosion morphology}
\end{figure*}

\begin{figure*}[htbp]%
	\centering
	\includegraphics[width=1.0\textwidth]{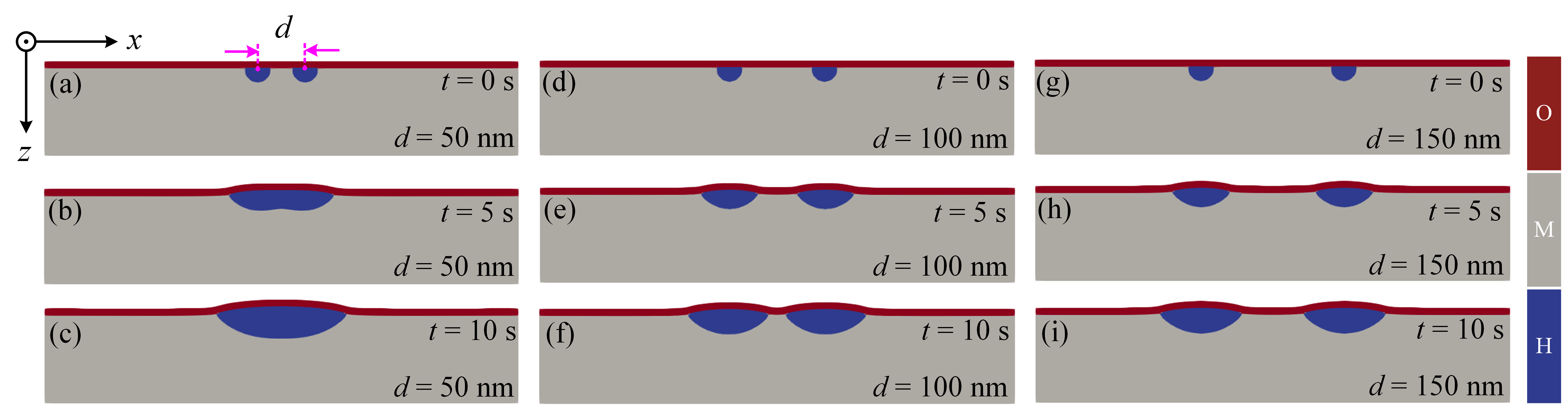}
	\caption{\label{figmultiple_spots_distance}Effect of the distance between corrosion spots on the evolution of corrosion morphology}
\end{figure*}

Fig. \ref{figmultiple_spots_number} displays the influence of the number $N$ of corrosion spots on the evolution of corrosion morphology, showing that the more the number of corrosion spots, the larger and flatter the corrosion area formed by the merging of corrosion spots. Fig. \ref{figmultiple_spots_distance} illustrates the impact of the distance $d$ between corrosion spots on the evolution of corrosion morphology. It is observed that as the distance between corrosion spots increases, the time required for interaction to occur also lengthens, indicating a delay in the merging process.

\begin{figure*}[htbp]%
	\centering
	\includegraphics[width=0.7\textwidth]{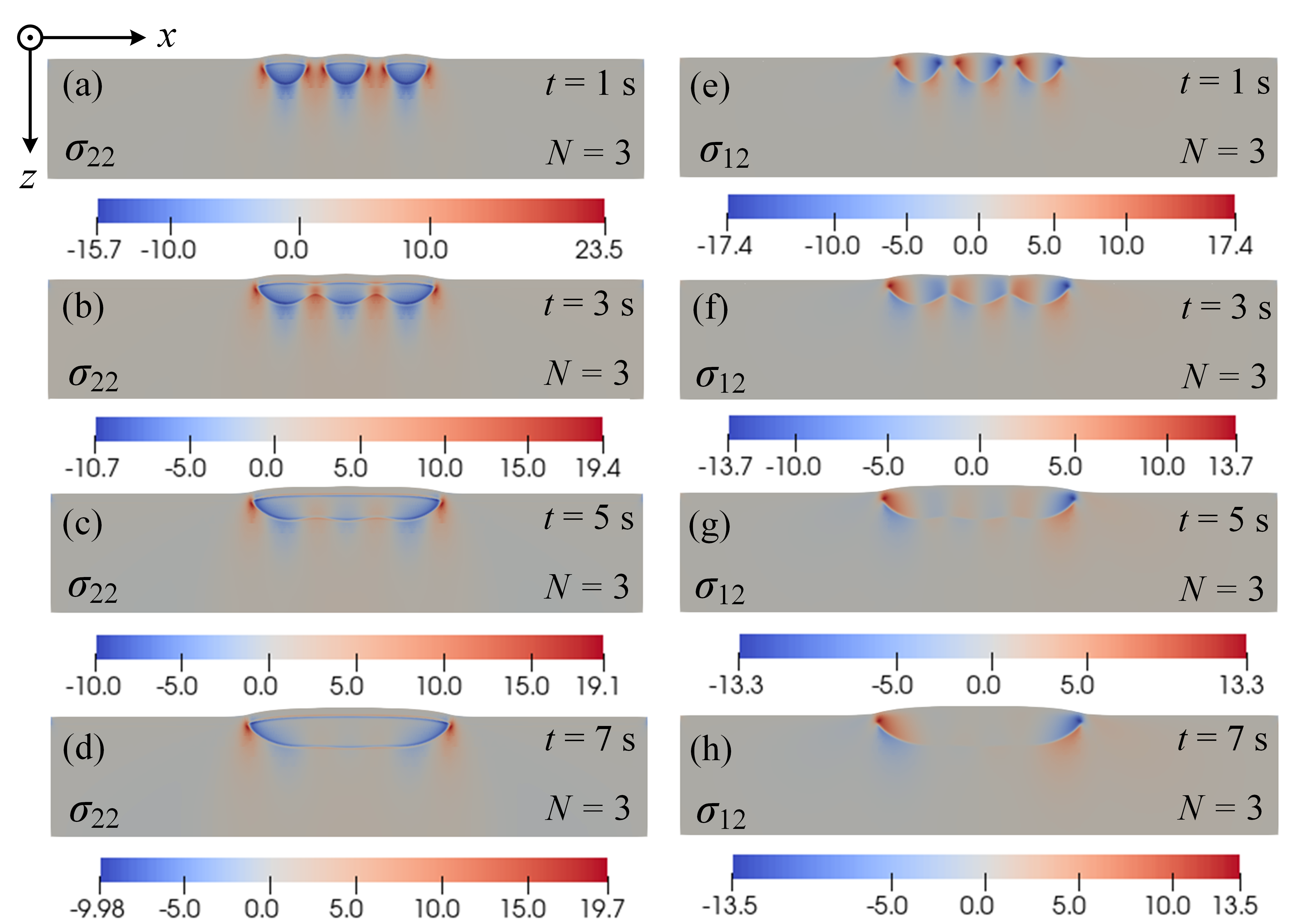}
	\caption{\label{figmultiple_spots_stress22and12}Effect of the merging of corrosion spots on the stresses $\sigma_{22}$ and $\sigma_{12}$}
\end{figure*}

In addition, Figs. \ref{figmultiple_spots_stress22and12}(a)-(d) and (e)-(h) show the effect of the merging of corrosion spots on the stresses $\sigma_{22}$ and $\sigma_{12}$, respectively. The merging of corrosion spots can also release the stresses $\sigma_{22}$ and $\sigma_{12}$ between adjacent corrosion points .

\subsection{Effect of the grain orientation on the corrosion degree and stresses $\sigma_{22}$ and $\sigma_{12}$}\label{sec_grain_orientation}

%Fig. \ref{figsinglecrystal_corrosion_degree_orientation} illustrates the impact of the grain orientation on the corrosion degree $(V_{h}/V)$, showing that the corrosion degree decreases with the increase of grain rotation angle $\theta_{R}$.  

\begin{figure*}[htbp]%
	\centering
	\includegraphics[width=0.7\textwidth]{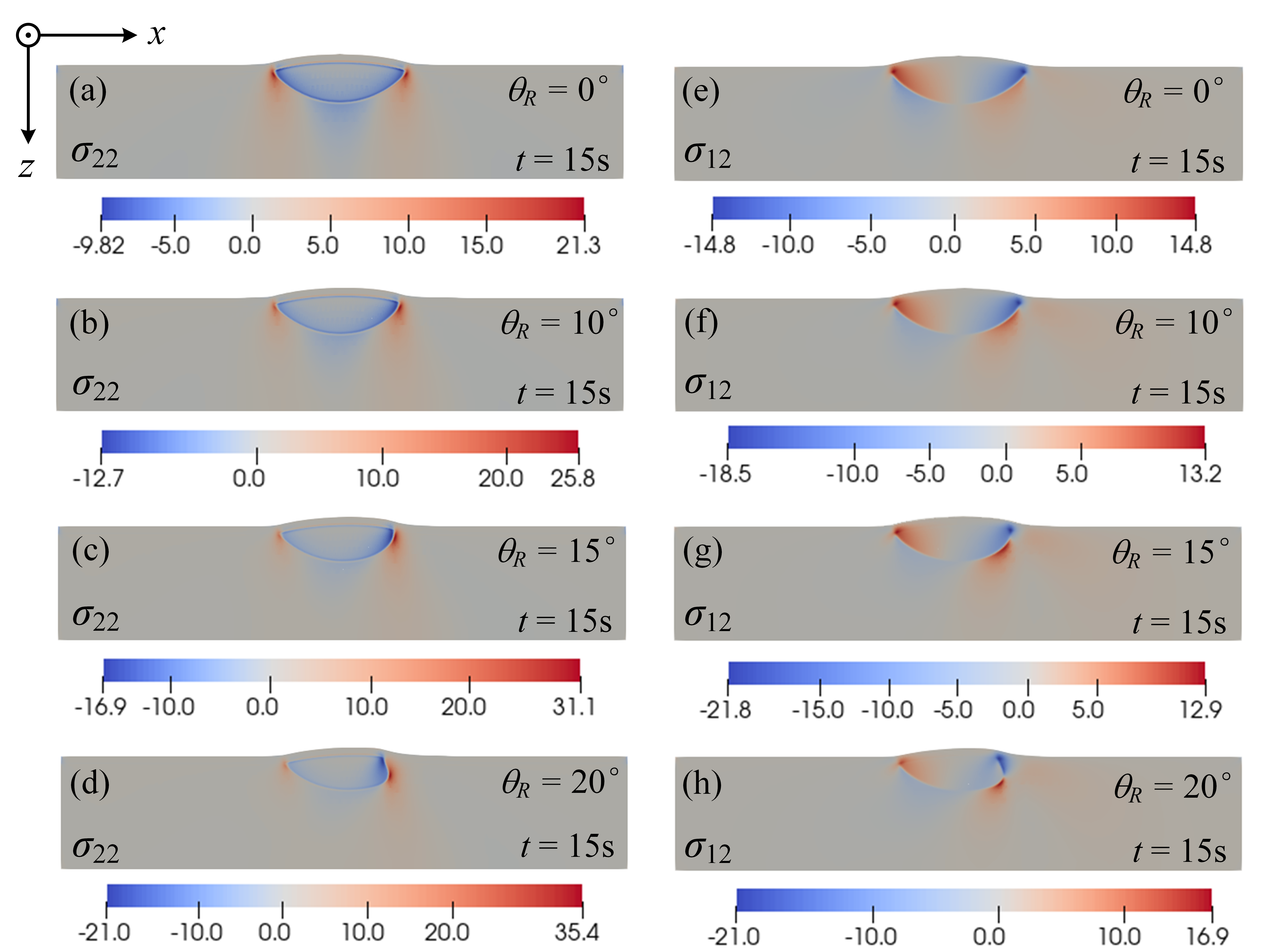}
	\caption{\label{figsinglecrystal_grain_orientation_stress22and12}Effect of the grain orientation on the stresses $\sigma_{22}$ and $\sigma_{12}$}
\end{figure*}

Figs. \ref{figsinglecrystal_grain_orientation_stress22and12}(a)-(d) and (e)-(h) respectively demonstrate the influence of the grain orientation on the stresses $\sigma_{22}$ and $\sigma_{12}$, showing that grain rotation also changes the symmetrical distribution of stresses $\sigma_{22}$ and $\sigma_{12}$. In particular, asymmetric shear stress $\sigma_{12}$ can cause the rotation of the hydride precipitate inside the metal, which may lead to torsional fracture of the surface oxide film.

%\begin{figure*}[htbp]%
%	\centering
%	\includegraphics[width=0.5\textwidth]{singlecrystal_corrosion_degree_orientation.png}
%	\caption{\label{figsinglecrystal_corrosion_degree_orientation}. Effect of the grain orientation on the corrosion degree}
%\end{figure*}

\section{Some additional researches on the Bicrystal}

\subsection{Effect of grain boundary on the stresses $\sigma_{22}$ and $\sigma_{12}$}\label{sec_GB} 
\begin{figure*}[htbp]%
	\centering
	\includegraphics[width=0.7\textwidth]{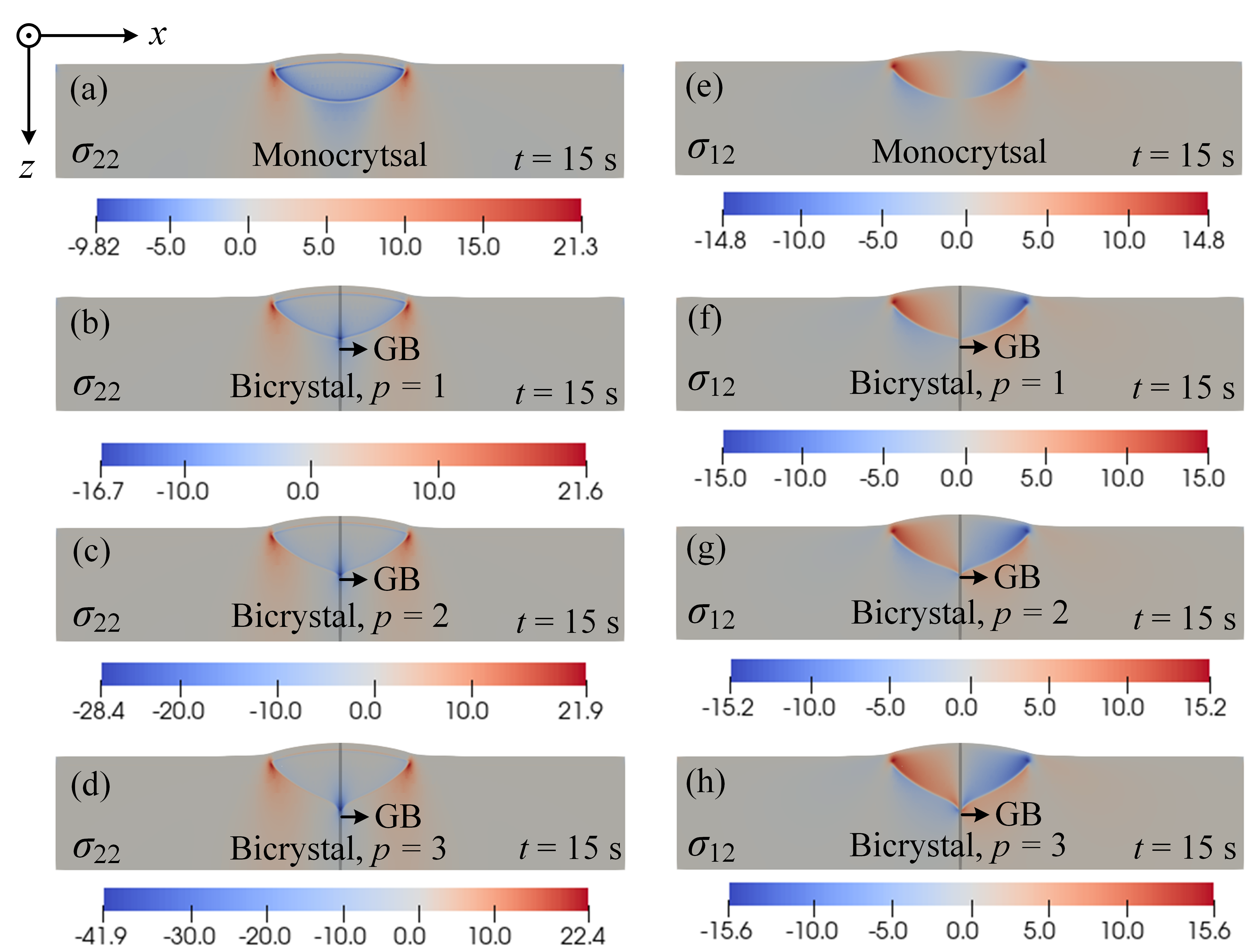}
	\caption{\label{figGB_and_p_parameter_stress22and12}Effect of the grain boundary on the stresses $\sigma_{22}$ and $\sigma_{12}$}
\end{figure*}

Figs. \ref{figGB_and_p_parameter_stress22and12}(a)-(d) and (e)-(h) show the effect of grain boundary on the stresses $\sigma_{22}$ and $\sigma_{12}$, respectively. We find that a compressive stress $\sigma_{22}$ concentrates in the grain boundary of bicrystal, but the shear stress distribution in the bicrystal is almost the same as that in the monocrystal.

\subsection{Effect of the angles between the grain boundary and metal surface on the corrosion morphology} \label{sec_theta_GB} 

In metalworking processes, the grain boundaries in the metal can form various angles ($\theta_{\mathrm{GB}}$) with the cutting plane, depending on the machining direction. The orientation of these angles has been observed to have an impact on hydride growth\cite{banos2016effect,banos2018review} and nucleation\cite{scott2007ud3} behavior. Our multiphase-field model is employed to investigate the influence of these angles on the evolution of corrosion morphology and stress during spot corrosion. Understanding and predicting the spot corrosion behavior on the metal surface at different cutting planes can provide valuable insights for practical applications. We set the angles of the two grains in the bicrystal as $0^{\circ}$ and $180^{\circ}$, respectively\cite{heo2019phase}.

\begin{figure*}[htbp]%
	\centering
	\includegraphics[width=0.7\textwidth]{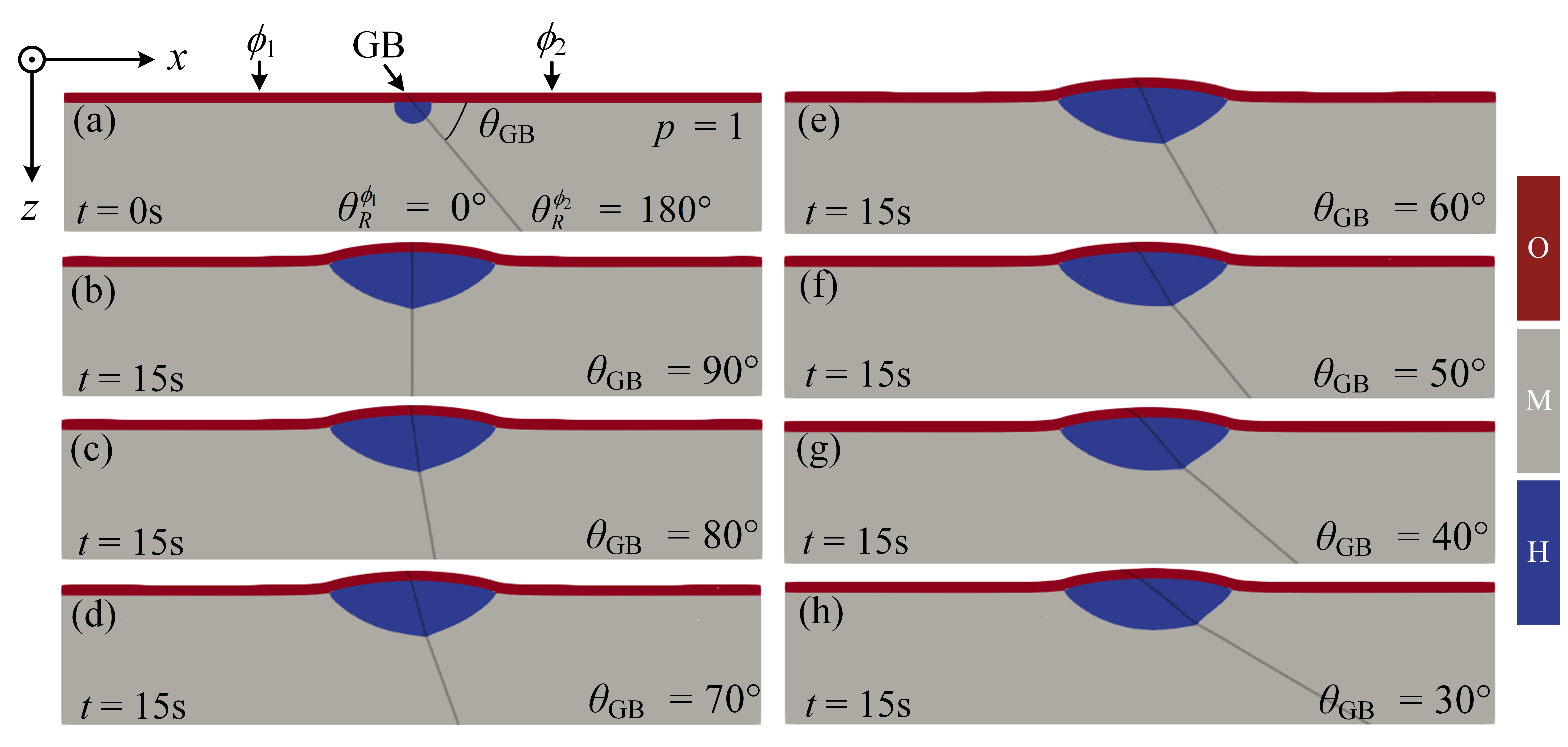}
	\caption{\label{figbicrystal_grain_angle_morphology}(a)-(h) Effect of the angles ($\theta_{\mathrm{GB}}$) between the grain boundary and metal surface on the corrosion morphology.}
\end{figure*}

\begin{figure*}[ht]%
	\centering
	\includegraphics[width=1.05\textwidth]{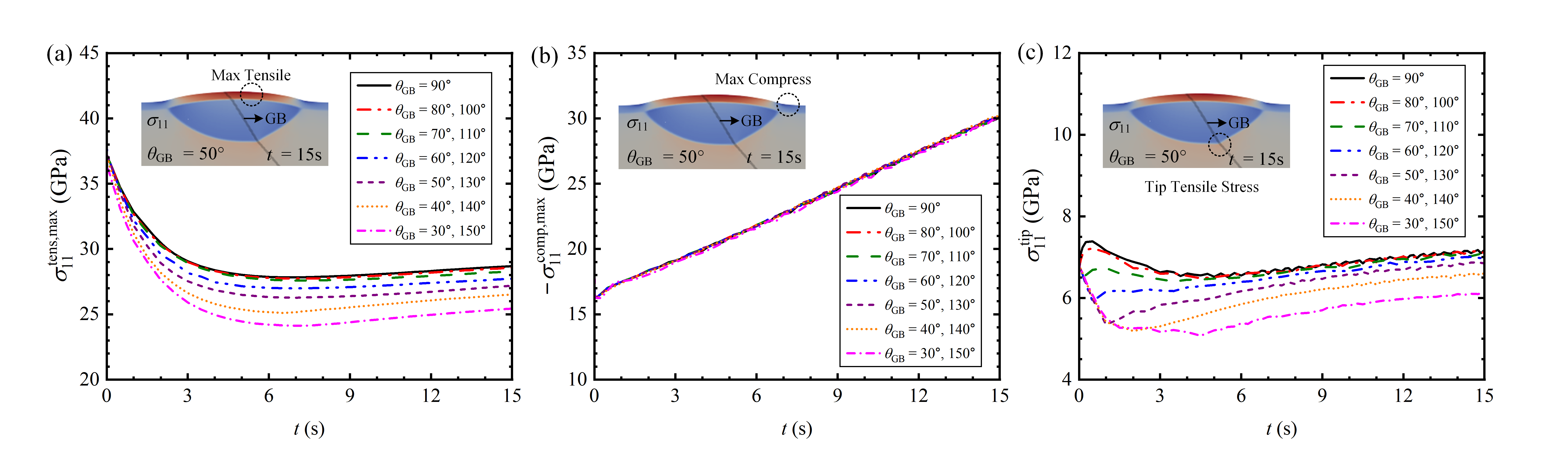}
	\caption{\label{figbicrystal_grain_angle_stress}Effect of different $\theta_{\mathrm{GB}}$ on the maximum (a) tensile and (b) compressive stresses $\sigma_{11}$ (GPa) in the oxide film; (c) Effect of different $\theta_{\mathrm{GB}}$ on the tip tensile stress in the metal matrix. }
\end{figure*}

Fig. \ref{figbicrystal_grain_angle_morphology} demonstrates the effect of the angles $\theta_{\mathrm{GB}}$ on the corrosion morphology when $t=15$ s.  It is evident from the figure that the variation in the angle $\theta_{\mathrm{GB}}$ has a impact on the position of the ``tip'' morphology. Figs. \ref{figbicrystal_grain_angle_stress}(a) and (b) show the influence of the angle $\theta_{\mathrm{GB}}$ on the maximum tensile and compressive stresses within the oxide film. Fig. \ref{figbicrystal_grain_angle_stress}(c) displays the influence of the angle $\theta_{\mathrm{GB}}$ on the tip tensile stress in the metal. A low angle $\theta_{\mathrm{GB}}$ during machining can reduce the maximum tensile stress in the oxide film and the tip tensile stress in the metal matrix during the corrosion process. However, the effect of the angle $\theta_{\mathrm{GB}}$ on the maximum compressive stress is almost negligible. This negligible effect may be due to the fact that the angle $\theta_{\mathrm{GB}}$ primarily affects the ``tip'' morphology at the grain boundary locally and has little impact on the overall hydride morphology.

\section{Some additional researches on the Polycrystal}\label{polycrystal}

\subsection{Effect of grain orientation variation on the corrosion morphology and material's stresses in the polycrystal system}

\begin{figure*}[ht]%
	\centering
	\includegraphics[width=0.6\textwidth]{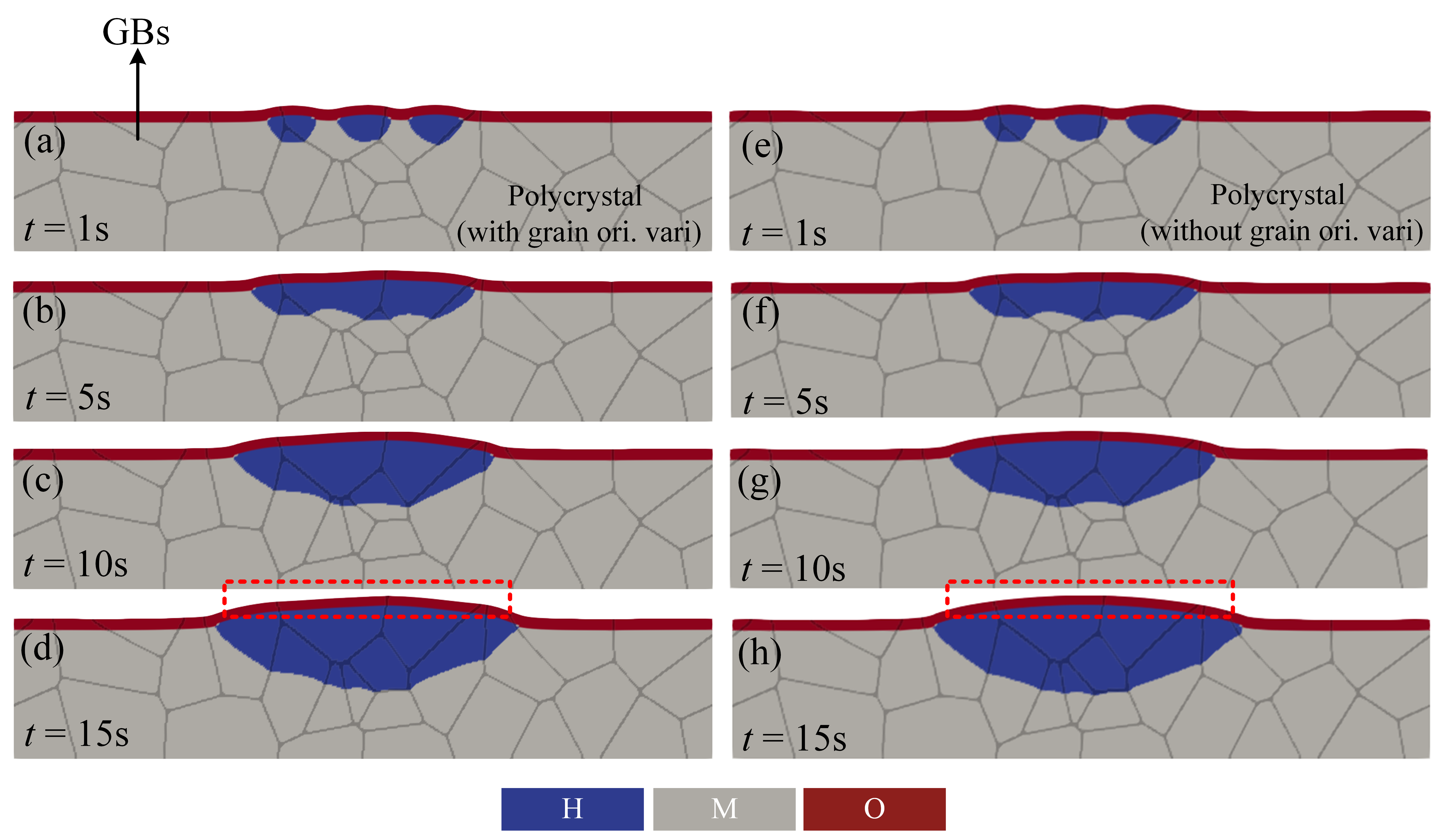}
	\caption{\label{figpolycrystal_morphology}(a)-(d) Evolution of morphology in the polycrystal uranium metal with different grain orientation variation. (e)-(h) Evolution of morphology in the polycrystal uranium metal without grain orientation variation.}
\end{figure*}

\begin{figure*}[htbp]%
	\centering
	\includegraphics[width=0.6\textwidth]{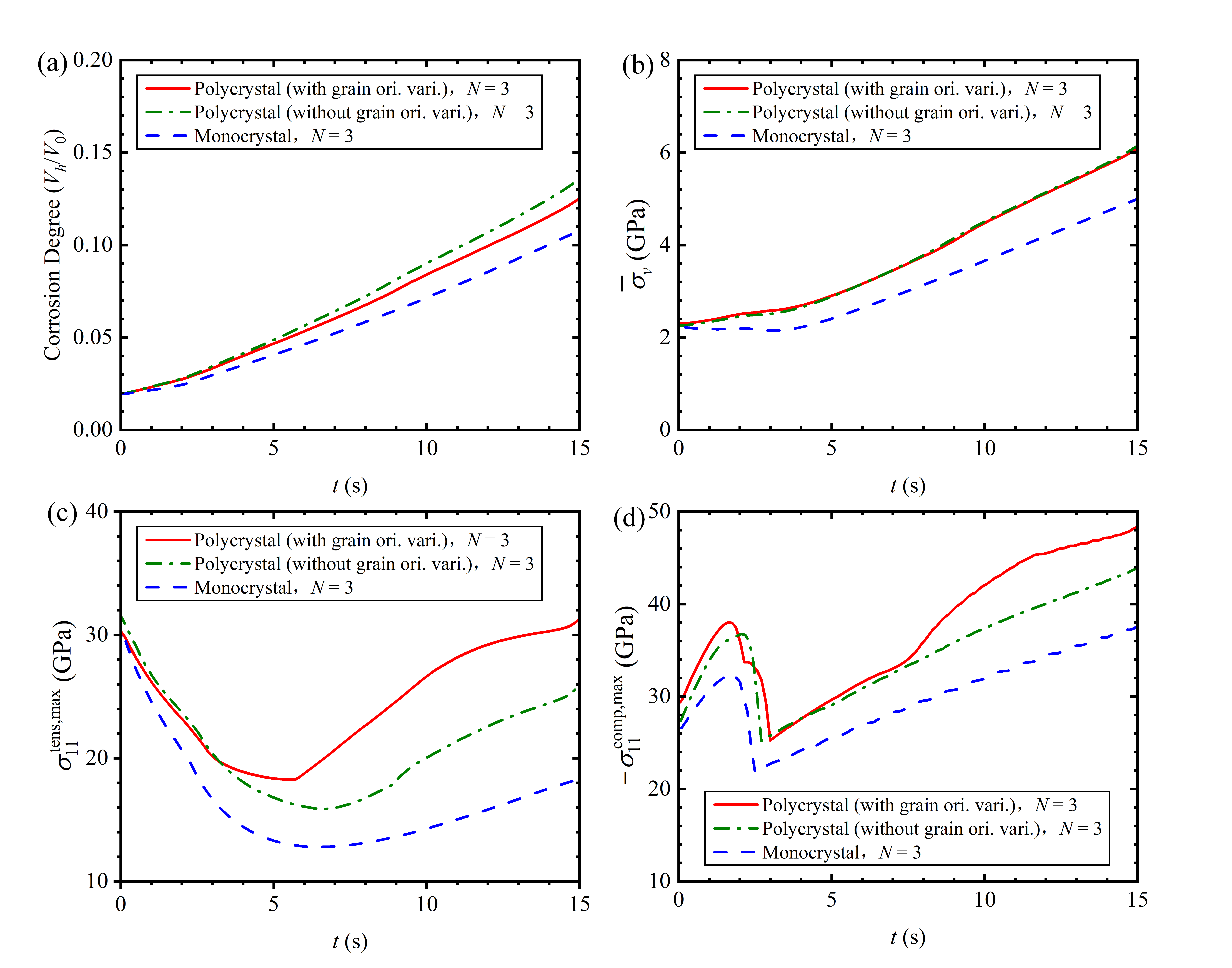}
	\caption{\label{figpolycrystal_stress_statistics}Corrosion and stress curves during the multi-spot corrosion process in the polycrystal with and without grain orientation variation: (a) Variation of the corrosion degree $(V_{h}/V_0)$ of the metal matrix versus time $t$; (b) Variation of the average von Mises stress $\bar{\sigma}_v$ of the materials versus time $t$; Variation of the maximum (c) tensile  ($\sigma^{\mathrm{tens,max}}_{11}>0$) and (d) compressive ($\sigma^{\mathrm{comp,max}}_{11}<0$) stresses versus time $t$ in the oxide film.} 
\end{figure*}

Figs. \ref{figpolycrystal_morphology}(a)-(d) and (e)-(h) demonstrate that the evolution of multi-spot corrosion morphology in the polycrystal with and without grain orientation variation, respectively. It can be seen that the corrosion morphology of polycrysal with the grain orientation variation is more irregular than that of polycrysal without the grain orientation variation. In addition, the grain orientation variation also makes the bulge of oxide film above the hydride irregular, which affects the stress distribution in the oxide film.

Fig. \ref{figpolycrystal_stress_statistics} shows the corrosion degree and stress curves during the multi-spot corrosion process in the polycrystal with and without grain orientation variation and monocrystal. The corrosion degree and average von Mises stress curves of the two polycrystalline cases are very similar. However, the variation of the maximum tensile and compressive stress of the oxide film is quite different in the two polycrystalline cases, which is due to the irregularity of the bulge of oxide film in the polycrystal with the grain orientation variation.

\section{Qualitative comparison between phase-field simulations and experimentally observed  corrosion morphology in the polycrystalline system}\label{sec_comparsion}
\begin{figure*}[htbp]%
	\centering
	\includegraphics[width=0.8\textwidth]{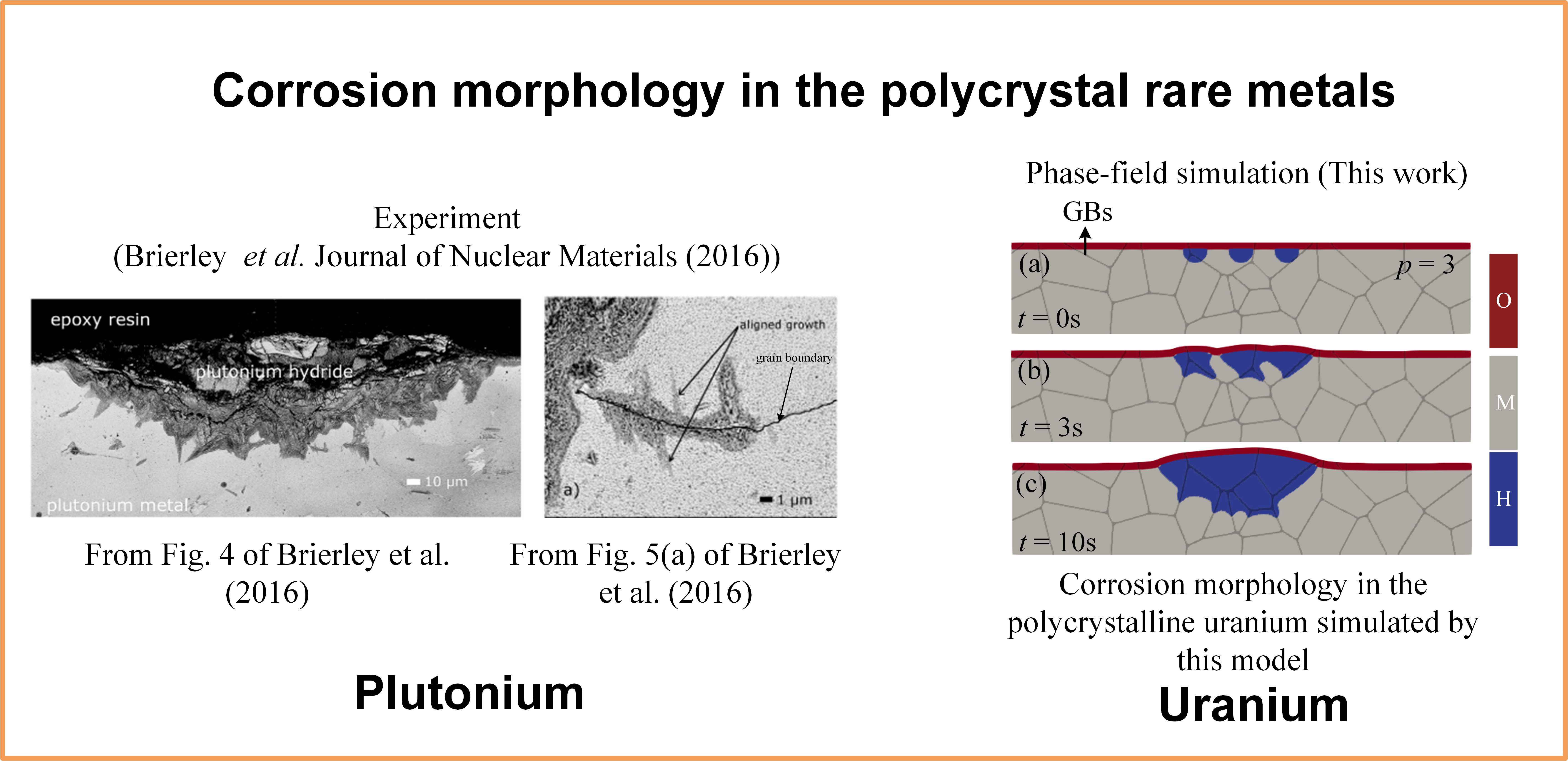}
	\caption{\label{fig_polycrystal morphology and experiment}Corrosion morphology in the polycrystal rare metals: Comparison between experimental (by Brierley et al.\cite{brierley2016anisotropic}) corrosion morphology in the polycrystalline plutonium and simulated (this work) corrosion morphology in the polycrystalline uranium.}
\end{figure*}

Our multiphase-field model qualitatively captures the corrosion morphology in the polycrystalline systems. We give a comparison between experimental and simulated corrosion morphology. Unfortunately, we do not find any available images of hydrogen-induced spot corrosion morphology about the polycrystalline uranium. Instead, the hydrogen-induced spot corrosion morphology (from Brierley et al.\cite{brierley2016anisotropic}) of polycrystalline plutonium with similar properties to uranium is shown in Fig. \ref{fig_polycrystal morphology and experiment}.

\bibliography{main_PF2_and_SI}

\clearpage